\documentclass[twocolumn]{aa} 
\usepackage[english]{babel}
\usepackage{graphicx}
\usepackage[varg]{txfonts}
\usepackage{natbib}
\usepackage[version-1-compatibility]{siunitx}
\DeclareSIUnit{\solm}{\, M_\odot}
\newcommand{\norm}[1]{\left\lVert#1\right\rVert} 
\newcommand{\ramses}{\textsc{Ramses}}
\newcommand{\ramsesrt}{\textsc{Ramses-RT}}
\usepackage{hyperref}
\bibpunct{(}{)}{;}{a}{}{,} 
\usepackage{amsmath}
\usepackage{esdiff}
\usepackage{xcolor}

\begin{document}

   \title{A new hybrid radiative transfer method for massive star formation}


   \author{R. Mignon-Risse
          \inst{1}
          \and
          M. Gonz\'alez \inst{1}
          \and
          B. Commer\c con \inst{2}
          \and
          J. Rosdahl \inst{3}
          }

   \institute{AIM, CEA, CNRS, Universit\'e Paris-Saclay, Universit\'e Paris Diderot, 
   			Sorbonne Paris Cit\'e, F-91191 Gif-sur-Yvette, France \\
         \email{raphael.mignon-risse@cea.fr}
         \and
             Centre de Recherche Astrophysique de Lyon UMR5574, ENS de Lyon, Univ. Lyon1, CNRS, Universit\'e de Lyon, 69007 Lyon, France
             \and
             Univ Lyon, Univ Lyon1, Ens de Lyon, CNRS, Centre de Recherche Astrophysique de Lyon UMR5574, F-69230, Saint-Genis-Laval, France
             }

   \date{Received: 30/08/2019 ; Accepted: 28/01/2020}

  \abstract
   {Frequency-dependent and hybrid approaches for the treatment of stellar irradiation are of primary importance in numerical simulations of massive star formation. 
    }
   {We seek to compare outflow and accretion mechanisms in star formation simulations. 
   We investigate the accuracy of a hybrid radiative transfer method using the gray M1 closure relation for proto-stellar
    irradiation and gray flux-limited diffusion (FLD) for photons 
    emitted everywhere else.}
   {We have coupled the FLD module of the adaptive-mesh refinement code \ramses{} with \ramsesrt{},
     which is based on the M1 closure relation and the reduced speed-of-light-approximation.
     Our hybrid (M1+FLD) method takes an average opacity at the stellar temperature for the M1 module, instead of the local environmental radiation field.
    Due to their construction, the opacities are consistent with the photon origin.
     We have tested this approach in radiative transfer tests of disks irradiated by a star for three levels of optical thickness and compared the temperature structure with the radiative transfer codes RADMC-3D and MCFOST.
     We applied it to a radiation-hydrodynamical simulation of massive star formation.}
   {Our tests validate our hybrid approach for determining the temperature structure of an irradiated disk in the optically-thin ($2\%$ maximal error) and moderately optically-thick (error smaller than $25 \%$) regimes.
   The most optically-thick test shows the limitation of our hybrid approach with a maximal error of $65\%$ in the disk mid-plane against $94\%$ with the FLD method. 
   The optically-thick setups highlight the ability of the hybrid method to partially capture the self-shielding in the disk while the FLD alone cannot. 
   The radiative acceleration is ${\approx}100$ times greater with the hybrid method than with the FLD.
   The hybrid method consistently leads to about $+50 \%$ more extended and wider-angle radiative outflows in the massive star formation simulation.
   We obtain a $17.6 \si{\solm}$ star at $t{\simeq}0.7 \tau_\mathrm{ff}$, while the accretion phase is still ongoing, with a mean accretion rate of ${\simeq}7 \times 10^{-4} \si{\solm.yr^{-1}}$.
   Finally, despite the use of refinement to resolve the radiative cavities, no Rayleigh-Taylor instability appears in our simulations, and we justify their absence by physical arguments based on the entropy gradient.}
   {}


   \maketitle
%

\section{Introduction}

    Massive stars shape the dynamical and chemical evolution of galaxies because of their powerful feedback in radiation, winds, explosions in supernova, and metal-enrichment. However, their formation remains a long-standing problem.
    Observationally, massive stars are embedded in dense clouds; they form on timescales much shorter than their low-mass counterpart \citep{motte_earliest_2007} and are likely to be located at distances larger than \SI{1}{kpc} from us, which makes their formation process challenging to observe.
    Two major scenarios are under active studies: the competitive accretion model and the turbulent core accretion model. 
    In the competitive accretion model \citep{bonnell_massive_2004}, all stars form in clusters and stars located at the center of the gravitational potential gain more mass and eventually become massive stars, via accretion and possibly merging processes.
    In this scenario, the initial-mass function (IMF) is built-up naturally.
    On the other hand, the turbulent core model \citep{mckee_formation_2003} is an extension of the low-mass star formation scenario. A massive and turbulent prestellar core gravitationally collapses while fragmentation is limited by turbulent, radiative, and magnetic support (e.g. \citealt{commercon_collapse_2011}). 
    The IMF is therefore linked to the prestellar core mass function since more massive stars form in more massive cores.
    For reviews on theories of massive star formation, we refer the reader to \cite{beuther_formation_2007}, \cite{zinnecker_toward_2007}, \cite{tan_massive_2014}, and \cite{krumholz_notes_2017}.
    
    There is no consensus regarding the accretion process and a way to probe it is to study outflows.
    Indeed, magnetic outflows are often associated with accretion (\citealt{blandford_hydromagnetic_1982}, \citealt{pelletier_hydromagnetic_1992}) as they remove the angular momentum from an accretion disk and they do not require strong magnetic fields.
    It is then possible to study the physics of accretion via the outflow properties \citep{pudritz_disk_2006} and in particular the accretion rate from the outflow velocity \citep{pelletier_hydromagnetic_1992}.
    In the case of massive stars, they can act as a channel for radiation to escape.
    Moreover, accretion modes differ from one scenario to another.
    Disk accretion is more likely to occur in the turbulent core accretion model, with high accretion rates ($\mathrm{\dot{M}} {\sim}10^{-4}{-}10^{-3} \si{\solm.yr^{-1}}$, \citealt{mckee_formation_2003}).
    More chaotic accretion mechanisms are associated with the competitive accretion model or with models including accretion via filaments.
    Recent observations by \cite{goddi_accretion_2018} reveal signatures of outflows whose direction varies through time.
    If they are perpendicular to the accretion disk \citep{blandford_hydromagnetic_1982} this indicates that the plane of accretion changes with time, favoring accretion via filaments and competitive accretion.
    The study of outflows can indeed help to distinguishing the accretion modes and these questions highlight the need for realistic outflow models (magnetic and radiative) in numerical simulations.
    Regarding the radiative outflows, this means the use of a radiative transport method well-suited in the optically-thin regime.
    Current and past studies of massive star formation have mainly focused on its radiative aspects.
   In a 1D spherically-symmetric approach, the radiative force of a massive (proto)star is expected to counteract gravity up to the point where accretion is stopped as computed analytically  \citep{larson_formation_1971} and then numerically \citep{kuiper_circumventing_2010}.
   Their results showed that the highest mass reached was \SI{40}{\solm}.
   Two-dimensional and 3D numerical simulations have permitted the emergence of a new accretion mode, the "flashlight effect" \citep{yorke_formation_2002} which allows the radiation to escape freely in the poles while material is accreted through the disk (\citealp{krumholz_formation_2009}, \citealp{kuiper_circumventing_2010}, \citealp{rosen_unstable_2016}, \citealp{harries_radiation-hydrodynamical_2017}).
   
   Monte-Carlo approaches are often used for solving radiative transfer problems for their accuracy but they are particularly expensive and their computational time scales with the number of radiative sources.
   This justifies the use of fluid description models such as the flux-limited diffusion (FLD) and the M1 methods for radiation-hydrodynamics (RHD).
   The first RHD calculations relied on the FLD closure relation for its simplicity and advantageous computational cost.
   However, it is more suited for the optically-thick regime while the use of a flux-limiter corrects the propagation speed in the optically-thin regime \citep{levermore_flux-limited_1981}.
   In addition, the FLD method does not permit to capture shadows behind very optically-thick gas.
   In the context of massive star formation, the flashlight effect is due to the nonisotropic character of the radiation field because the optical thickness is very different in the disk direction and in the cavities direction.
   Therefore, numerical developments regarding radiative transfer have been made, especially in the optically-thin limit.
   Recent approaches treat stellar irradiation in a more consistent way with ray-tracing (\citealp{kuiper_fast_2010}, \citealt{kim_modeling_2017}), long-characteristics \citep{rosen_hybrid_2017}, Monte-Carlo radiative transfer (\citealt{haworth_radiation_2012}, \citealp{harries_radiation-hydrodynamical_2017}, who took advantage of the independency between photon packets to parallelize efficiently the Monte-Carlo step), and the M1 closure relation (\citealt{levermore_relating_1984}, \citealt{gonzalez_heracles:_2007}, \citealt{aubert_teyssier_2008}, \citealt{rosdahl_ramses-rt:_2013}, \citealt{kannan_arepo-rt_2019}, this work).
   
   Multi-dimensional simulations using the FLD approximation or hybrid approaches show stars with mass above the \SI{40}{\solm} limit obtained in 1D: with the FLD method only, \cite{yorke_formation_2002} and \cite{krumholz_formation_2009} form a star of \SI{{\approx}42}{\solm} from a \SI{120}{\solm} and \SI{100}{\solm} prestellar core, respectively.
   The additional treatment of direct irradiation in hybrid approaches has been shown not to impact the stellar mass significantly: \cite{klassen_simulating_2016} have obtained stars as massive as \SI{43.7}{\solm} from an initial mass of \SI{100}{\solm}, the simulations of \cite{rosen_unstable_2016} show a \SI{40}{\solm} star and \cite{kuiper_circumventing_2010} obtain a \SI{56.5}{\solm} star from a \SI{120}{\solm} prestellar core in several free-fall times.
    Most of these works put lower-limit on the stellar mass because the accretion phase is not finished yet at the end of the run (except for \citealp{kuiper_circumventing_2010}).
    These investigations have also noted the formation of polar cavities dominated by the stellar radiative pressure, enhanced by the particular treatment of stellar irradiation.
    
    In addition, \cite{krumholz_formation_2009} and \cite{rosen_unstable_2016} observe the onset of radiative Rayleigh-Taylor instabilities in the radiation-pressure-dominated cavities that feed the star-disk system and help accreting mass onto the central star via the flashlight effect.
    However, these instabilities have not been observed in the work of \cite{kuiper_circumventing_2010} and \cite{klassen_simulating_2016}.
    \cite{kuiper_stability_2012} argue that a gray FLD model, as used in \cite{krumholz_formation_2009}, underestimates the radiation force in the cavity and can artificially lead to the apparition of these instabilities.
    With a frequency-dependent hybrid model and a Cartesian grid, \cite{klassen_simulating_2016} did not obtain such instabilities while \cite{rosen_unstable_2016} did.
    The difference in their results can be explained by the use of refinement by \cite{rosen_unstable_2016} to resolve the seeds of the instabilities, the smallest modes being the more unstable \citep{jacquet_radiative_2011}.
    Contrarily, the spherical grid without additional refinement in the cavities used by \cite{kuiper_stability_2012} would not permit to refine them.
    It is thus unclear yet whether this mechanism is at work during the formation of massive stars.
    Meanwhile, it is clear that disk-accretion is sufficient to reach masses consistent with the massive stars observed.
    Our hybrid method, implemented in the Cartesian adaptive-mesh refinement (AMR) code \ramses{} \citep{teyssier_cosmological_2002} help us to establish the importance of these accretion mechanisms by capturing the nonisotropy of the radiation field.
    
    This paper is organized as follows.
    Section~\ref{sec:model} presents the equations of the flux-limited diffusion method and the M1 method, along with their coupling and implementation in the \ramses{} code.
    We present the tests and the validation of our hybrid approach in Section~\ref{sec:numtest}
    and its application to the collapse of a massive prestellar core leading to the formation of a massive star in Section~\ref{sec:col}.
    We discuss our results in Section~\ref{sec:ccl}

\section{Methods}
\label{sec:model}

In this section, we present the equations of the hybrid radiative transfer approach and its implementation: the M1 method for the stellar irradiation and the flux-limited diffusion method for the dust emission.

	\subsection{Coupling flux-limited diffusion and M1}
   The FLD method (\citealt{levermore_flux-limited_1981}) and the M1 method \citep{levermore_relating_1984} are fluid descriptions of the radiation field.
   They are based on moments
   of the equation of radiative transfer, that is, the equation of conservation of the radiation
    specific intensity $I_\nu(\boldsymbol{x},t; \boldsymbol{n})$
   with the propagation, the absorption and the emission \citep{mihalas_foundations_1984}
   \begin{equation}
   \frac{1}{\mathrm{c}} \diffp{I_\nu}{t} + \boldsymbol{n} \cdot \nabla I_\nu
   = - \kappa_\nu \rho I_\nu + \eta_\nu.
   \label{eq:rt}
   \end{equation}
   Here, $I_\nu(\boldsymbol{x}, t; \boldsymbol{n})$ is the amount of energy of a photon beam
   at a given position $\boldsymbol{x}$ and time $t$, in direction $\boldsymbol{n}$ and 
   per unit frequency.
   $\mathrm{c}$ is the speed of light, $\kappa_\nu$ is the extinction coefficient (absorption and scattering contributions), $\rho$ is the local density and
   $\eta_\nu$ is the emission coefficient. 
   The sum of dust and gas contributions to the medium opacity weighted with the dust-to-gas ratio are encapsuled in $\kappa_\nu$.
   We assume local thermodynamical equilibrium (LTE) and do not consider scattering (see the justification in Appendix~\ref{app:isoscat}), hence the emission coefficient $\eta_\nu$ is
   a source function proportional to the Planck function $B_\nu(T)$ and 
   the extinction coefficient $\kappa_\nu$ is just an absorption coefficient.
   Equation~(\ref{eq:rt}) must be solved at each hydrodynamical timestep to evolve the radiation field, which still depends on six variables ($\boldsymbol{x},\boldsymbol{n}, \nu$).
   In addition, we want to couple it to hydrodynamics.
   This motivates the need for taking moments of the equation of radiative transfer.
   Hence we lose some of the angular information but it reduces the number of variables
   to four, at each timestep.
   The radiative energy density $E_\nu$, the radiative flux $\boldsymbol{F}_\nu$
   and the radiative pressure tensor $\mathbb{P}_\nu$ are defined as the $0$-th, $1$-st and $2$-nd
   moments of the radiative intensity $I_\nu$, respectively.
   
   Each system of moment equations involves the $i$-th and the $(i+1)$-th moments,
   hence we need a closure relation.
   The FLD scheme is based on the diffusion approximation, which is suited for high optical depths, when photons propagate in a random walk in the material (\textit{e.g.}, in stellar interiors).
   It has been commonly used as a first step to introduce radiation into hydrodynamical codes (\citealt{krumholz_equations_2007}, \citealt{kuiper_radiative_2008}, \citealt{tomida_radiation_2010}).
    In the FLD model, the equation to be solved is the equation of conservation of the radiative energy.
   Once integrated over all frequencies (often called a gray approximation) it gives
   \begin{equation}
    \diffp{E_{\mathrm{r}}}{t}   
   - \nabla \cdot \left( \frac{\mathrm{c} \lambda}
   	{\kappa_{\mathrm{R}} \, \rho } \nabla E_{\mathrm{r}} \right)
   = \kappa_{\mathrm{P}} \, \rho \mathrm{c} \left( \mathrm{a} T^4 -  E_{\mathrm{r}} \right),
   \label{eq:flde}
   \end{equation}
   where $E_\mathrm{r}$ is the frequency-integrated radiative energy, $\lambda$ is the flux-limiter and is built to recover the right propagation speed in optically-thin and -thick media \citep{levermore_flux-limited_1981}.
   The opacities are given by $\kappa_{\mathrm{P}}$ and  $\kappa_{\mathrm{R}}$, which are are Planck's and Rosseland's mean opacities, respectively.
   Thermal radiation is modeled as the Planck function $B(\nu,T)$, therefore
   under the gray model Planck's and Rosseland's opacities are respectively defined as
   \begin{equation}
   \kappa_\mathrm{P} = \frac{\int_{0}^{\infty} \kappa_\nu  B(\nu) d\nu}{\int_{0}^{\infty} B(\nu)  d\nu},
   \label{eq:kp}
   \end{equation}
   and
   \begin{equation}
   \kappa_\mathrm{R} = \frac{\int_{0}^{\infty} \diffp{B(\nu, T)}{T} d\nu}
   {\int_{0}^{\infty} \frac{1}{\kappa_\nu} \diffp{B(\nu, T)}{T}  d\nu},
   \label{eq:kr}
   \end{equation}
   where the temperature derivatives appear from the chain rule applied to $\nabla B$.
   The $\mathrm{a} T^4$ term in Eq.~\ref{eq:flde} arises from the integral of the Planck function over all frequencies, with $\mathrm{a}$ the radiation constant.
   We approximate the mean opacity of the radiative energy term as the Planck mean opacity.
   
   On the other hand, with the M1 method we take the zeroth and 
   first moments of the equation of radiative transfer \citep{levermore_relating_1984}.
   Within the M1 method we obtain the following system for the radiative energy and flux conservation, in the gray approximation as well
   \begin{equation}
   \begin{aligned}
   \diffp{E_\mathrm{r}}{t}  + \, \, \, \, \nabla \cdot \boldsymbol{F}_\mathrm{r}
   &= - \rho \kappa_\mathrm{P} \mathrm{c} E_\mathrm{r} + \dot{E}_\mathrm{r}^\star, \\
  \diffp{\boldsymbol{F}_\mathrm{r}}{t}  + \mathrm{c}^2 \nabla \cdot \mathbb{P}_\mathrm{r}
   &= -
   \rho \kappa_\mathrm{P} \mathrm{c} \boldsymbol{F}_\mathrm{r},
   \end{aligned}
   \label{eq:ef}
   \end{equation}   
   where $\dot{E}_\mathrm{r}^\star$ is the rate of radiative energy injected from stellar sources. $\boldsymbol{F}_\mathrm{r}$ and $\mathbb{P}_\mathrm{r}$ are the frequency-integrated radiative flux and pressure respectively.

   One main asset is that the directionality of the photons beam is well-retained.
   The M1 method is able to model shadows to some extent (see \citealp{gonzalez_heracles:_2007}), in an irradiated accretion disk for instance, while FLD is not.
   In addition, as a moment method the computing cost does not scale with the number of sources.

   Our goal is to take advantage of both methods, that is, the FLD method for an optically-thick medium and M1 for irradiation.
   Both FLD and M1 methods described above can involve several groups of photons or only one
   with frequency-averaged opacities.
   In our study however we restrict ourselves to one group of photons treated with each method.
   
   In massive star formation simulations the dynamical influence exerted 
   by the radiative feedback is of main importance
   as well as the thermal structure of the accretion flow (\textit{e.g.}, for fragmentation).
   However, doing so requires to retain to some extent the directionality of the photons emitted by the star to compute the direct radiative force.
   Breaking this isotropy is consistent with probing nonisotropic modes of accretion (disk or filaments).
   Secondly, it requires to distinguish the opacities between stellar photons, which have a UV-like energy and relatively high opacities,
   and photons emitted by the dust, which have a IR-like energy and relatively low opacities.
   
   Our method is to inject the stellar photons into the group of photons treated with the M1 scheme.
   The gray opacity used with the M1 corresponds to the Planck mean opacity at the stellar temperature, $\kappa_\mathrm{P}(T_\star)$, written $\kappa_\mathrm{P,\star}$ for the sake of readability.
   Once these photons are absorbed by the medium they are depleted from the M1 group as they heat the gas. The gas reemission is treated with the FLD method.
   In a first approach we do not deal with ionization states and leave this to further work.
   The set of equations that are to be solved are
   \begin{equation}
   \begin{aligned}
   \diffp{E_{\mathrm{M1}}}{t}  + \nabla \cdot \boldsymbol{F}_\mathrm{M1}
   &= - \kappa_\mathrm{P,\star} \, \rho \mathrm{c} E_\mathrm{M1} + \dot{E}_\mathrm{M1}^\star, \\
   \diffp{\boldsymbol{F}_\mathrm{M1}}{t}  + \mathrm{c}^2 \nabla \cdot \mathbb{P}_\mathrm{M1}
   &= - \kappa_\mathrm{P,\star} \, \rho \mathrm{c} \boldsymbol{F}_\mathrm{M1}, \\
  \diffp{E_{\mathrm{fld}}}{t} 
   - \nabla \cdot \left( \frac{c \lambda}
   	{\kappa_{\mathrm{R,fld}}} \nabla E_{\mathrm{fld}} \right)
   &= \kappa_{\mathrm{P,fld}} \, \rho \mathrm{c} \left( \mathrm{a} T^4 - E_{\mathrm{fld}} \right), \\
   C_\mathrm{v} \diffp{T}{t} 
   &= \kappa_\mathrm{P,\star} \, \rho \mathrm{c} E_\mathrm{M1}
   + \kappa_{\mathrm{P,fld}} \, \rho \mathrm{c} \left(E_{\mathrm{fld}} - \mathrm{a} T^4  \right),
   \end{aligned}
   \end{equation}
   where $\dot{E}_\mathrm{M1}^\star$ is the stellar radiation injection term, 
   and $\kappa_\mathrm{P,\star}\rho  \mathrm{c} E_\mathrm{M1}$ couples the M1 and the FLD methods via the equation of evolution of the internal energy.
   We use the ideal gas relation for the internal specific energy $\epsilon = C_\mathrm{v} T$
   where $C_\mathrm{v}$ is the specific heat capacity at constant volume.
   This equation closes the system and is used to evolve the gas temperature together with the radiative quantities.
   
   \subsection{Radiative acceleration}
   \label{subs:radacc}
   
   In addition of improving the thermal coupling between stellar irradiation and gas, our implementation is meant to affect the gas dynamics via a more accurate and less isotropic approach for the radiative acceleration than the FLD approximation thanks to the equation of evolution of the stellar radiative flux. We are interested in comparing the radiative acceleration with the hybrid method and with the pure FLD method.
   
	In the frame of RHD (for the full expression of RHD equations we refer the reader to \citealp{mihalas_foundations_1984}), the radiative acceleration at a given frequency is equal to 
    \begin{equation}
    \boldsymbol{a}_{\mathrm{rad},\nu} = \kappa_\nu \boldsymbol{F}_\nu / \mathrm{c}.
    \end{equation}

    However, gray FLD and M1 methods do not share the same expression for the radiative flux.
    On one hand, the M1 model includes the Planck mean opacity as the flux-averaged opacity --- which means that momentum is transferred each time a photon is absorbed --- so the radiative acceleration is given by
    \begin{equation}
    \boldsymbol{a}_{\mathrm{rad,M1}} = \kappa_\mathrm{P,\star} \boldsymbol{F}_{\mathrm{M1}} / \mathrm{c}.
    \end{equation} 

    On the other hand, the gray radiative acceleration in the FLD approximation is given by
    \begin{equation}
    \boldsymbol{a}_{\mathrm{rad,fld}} = - \frac{\lambda}{\rho} \nabla E_\mathrm{fld}.
    \end{equation}
    
    Taking the asymptotic values of $\lambda$ into account (see \citealt{levermore_relating_1984}), we get $\norm{\boldsymbol{a}_{\mathrm{rad,thin}}} =  \kappa_\mathrm{R} E_\mathrm{fld}$ in the limits of low optical depth and $\boldsymbol{a}_{\mathrm{rad,thick}} =  \frac{1}{3 \rho} \nabla E_\mathrm{fld}$ for high optical depth.
   
   We recall that $\kappa_\mathrm{R}$ is a harmonic mean which favors low absorption bands while $\kappa_\mathrm{P}$ is an arithmetic mean which favors high absorption bands.
   As a consequence, we expect a higher radiative acceleration with the hybrid method than with the FLD method.
   
  \subsection{Implementation}
  \label{subs:methods}
  
     The \ramses{} code (\citealt{teyssier_cosmological_2002}) is a 3D adaptive-mesh refinement Eulerian code.
   We use a version of \ramses{} which has been widely used for star formation simulations (\citealt{commercon_collapse_2011,commercon_synthetic_2012}, \citealt{joos_protostellar_2012}, \citealt{hennebelle_magnetically_2016}, \citealt{vaytet_protostellar_2018}).
   
   The hydrodynamical solver of \ramses{} relies on finite volume methods (variables are volume-averaged over the cell) and a second-order Godunov method is used to evolve hydrodynamical variables.
   This code includes 
   the FLD (\citealt{commercon_radiation_2011}, C11 hereafter) and the M1 method within \ramsesrt{} (\citealt{rosdahl_ramses-rt:_2013}\footnote{They express quantities in terms of photon number densities and consider several photo-absorbing species ($\ion{H}{I} , \ion{He}{I}$ and $\ion{He}{II}$) while we focus on a dust-and-gas mixture.}, R13 hereafter),
   both coupled to the hydrodynamics.
   For the FLD method, a time-splitting approach is performed.
   A predictor-corrector MUSCL scheme is used, where
   the predictor step is made under the diffusion approximation so the radiative pressure is isotropic and nonisotropy is taken into account in the corrector step.
   The hyperbolic part of the FLD solver relies on the second-order Godunov scheme of \ramses{} and fluxes are estimated with an approximate Riemann solver (Lax-Friedrich, HLL, HLLD, etc.).
   The diffusion and radiation-matter coupling are handled in the implicit part of the time-splitting scheme.
   The diffusion part of the FLD solver is second-order accurate in space.
   The M1 module estimates fluxes with a Riemann solver (HLL or GLF).
   In this work we use the GLF solver because it captures better the isotropy of stellar radiation than HLL, and the reduced flux approximation for the direct radiative force (see Appendix B of \citealp{rosdahl_galaxies_2015} and discussion in \citealt{hopkins_rp_2019}).
   
   The radiative transfer puts a heavy constraint on the timestep because
   the Courant-Friedrich-Lewy (CFL) condition forbids the propagation of signals
   through more than one cell in one timestep, for explicit schemes.
   For the hydrodynamics this speed is the sound speed but for radiative transfer
   this is the speed of light and this would impose a timestep ${\sim}1000$ times shorter.
   Therefore, both C11 with the FLD method and R13 with the M1 scheme have used a workaround.
   
   On one hand, the FLD solver for diffusion and radiation-matter coupling is implicit and therefore is unconditionally stable.
   The system composed of the internal energy and the radiative energy equations in their discretized form
    leads to the inversion of a matrix computed with the conjugate gradient or biconjugate gradient algorithm, in the case of multigroup radiative transfer \citep{gonzalez_multigroup_2015}.
    The number of iterations to converge scales with the number of cells.
    We note that the isotropic radiative pressure contributes to the total pressure in the explicit solver of \ramses{}, which therefore must satisfy the CFL condition.
    As a consequence, the radiative pressure must be taken into account when computing the timestep allowed by the CFL condition \citep{commercon_radiation_2011}.
    
   On the other hand, the M1 solver is fully explicit with a first-order Godunov scheme, thus it obeys the CFL condition.
   The trick used here is the reduced speed-of-light approximation \citep{gnedin_multi-dimensional_2001}.
   In this approximation, the propagation of light is not restricted by the speed of light but by the speed
   of the fastest wave, which is the speed of ionization front in the original paper and the fastest hydrodynamical speed in our case.
   An additional subcycling method relaxes this constraint.
   This leads to a timestep set by the hydrodynamical CFL condition.
   
   The injection of energy from the stellar source into the M1 photons is made via a sink algorithm \citep{bleuler_towards_2014}.
   In this work, we retrict ourselves to one stellar source for the M1 photons.
   The M1 module ensures the propagation and absorption of the stellar radiation while the FLD module deals with the heating by the stellar radiation and treats the reemission.
   We test the accuracy of the hybrid approach with respect to the FLD module alone, since it was used in previous massive star formation calculations with the \ramses{} code (\citealp{commercon_collapse_2011}).

   \section{Numerical tests}  
   \label{sec:numtest}
   
We test the hybrid method in a pure radiative transfer case (\textit{i.e.}, no hydrodynamics): a static disk irradiated by a star.
We compare the temperature structure obtained with results from Monte-Carlo radiative transfer codes.
We explore three levels of optical thickness integrated along the disk mid-plane: $\tau=0.1$, $\tau=100$ and $\tau=10^3$.
The parameters and results are summarized in Table~\ref{table:tests}.
We refer the reader to Appendix~\ref{app:perform} for performance tests.
  
\begin{figure}
  \resizebox{\hsize}{!}{\includegraphics{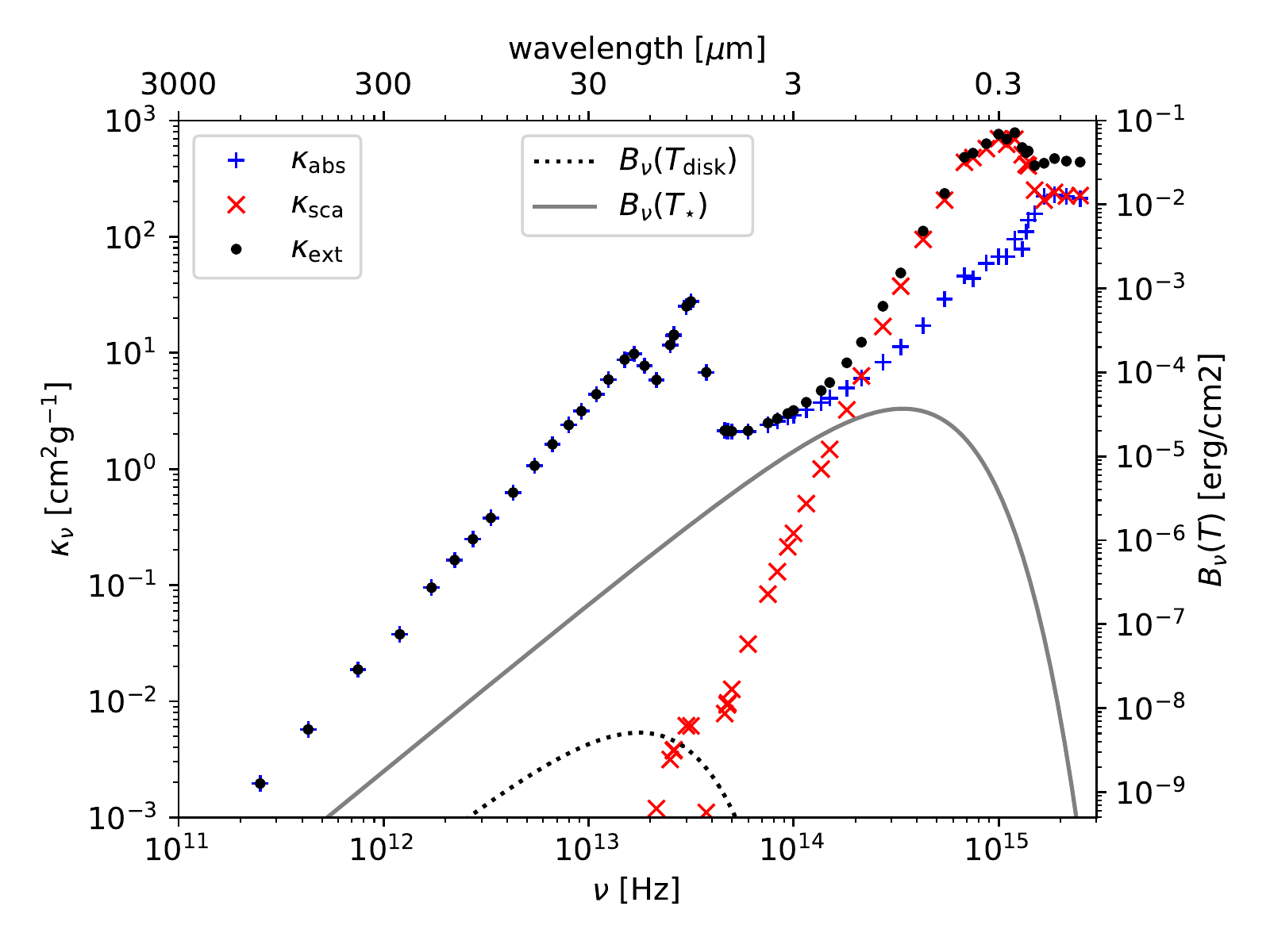}}
  \caption{Frequency-dependent opacities and blackbody spectra for $T_\mathrm{disk}=300$~K and $T_\star=5800$~K. Opacities are absorption (blue pluses), scattering (red crosses) and extinction (black dots) coefficients for the dust-and-gas mixture used in the Pasccuci test. The table contains 61 frequency bins and data are taken from \cite{draine_optical_1984}. Apart from the broad opacity features at about $10$ and \SI{20}{\mu m}, which correspond to Si-O vibrational transitions, the opacity generally increases with the photon frequency. The opacity at stellar-like radiation frequencies is generally greater than at disk-like radiation frequencies.}
  \label{fig:opa_pasc}
\end{figure}

\begin{figure}
  \resizebox{\hsize}{!}{\includegraphics{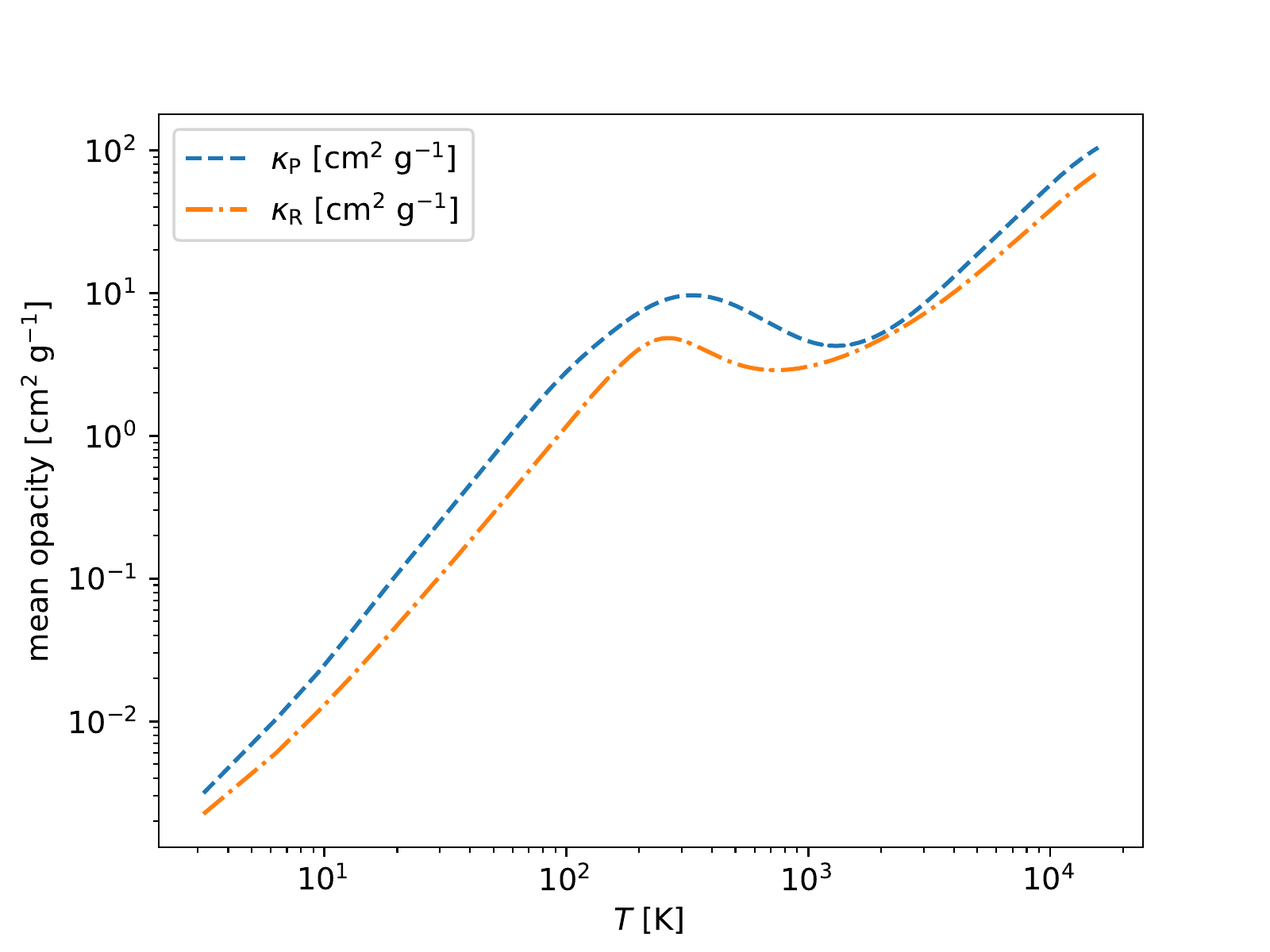}}
  \caption{Planck's (blue dashed curve) and Rosseland's (orange dot-dashed curve) mean opacities, as a function of temperature in the Pascucci setup.}
  \label{fig:kp_kr_pasc}
\end{figure}

\begin{table}
\scriptsize
\caption{Results from pure radiative transfer tests.}
\label{table:tests}    
\centering 
\begin{tabular}{l | l S[table-format=5] l S} 
    \hline
	 Ref. & $\tau$ & {$T_\star$ (K)} & Method & {$(\Delta T)_{\mathrm{max,r}}$ (\%)}  \\ 
	\hline
	\cite{pascucci_2d_2004} & $0.1$ & 5800  & FLD & 62  \\
	                		 &  	 & 		  & Hybrid & 2  \\
	                       & $0.1$  & 15000  & FLD & 65  \\
			                &  	    & 		 & Hybrid & 3  \\
	\hline
	\cite{pascucci_2d_2004} & $100$ & 5800  & FLD & 36  \\ 
			                & 		&        & Hybrid & 25  \\ 
	   & $100$ & 15000  & FLD & 57   \\
			 & 		 &	  & Hybrid & 31 \\
	\hline
	\cite{pinte_benchmark_2009} & $10^3$ & 4000  & FLD & 94  \\
			 & 	 	& 		 & Hybrid &  65 \\
	\hline
\end{tabular}
\end{table}

\subsection{Optically-thin and moderately optically-thick regimes: Pascucci's test}
\label{sec:pasc}

\subsubsection{Physical and numerical configurations}

\begin{figure*}
\centering
  \includegraphics[width=9cm]{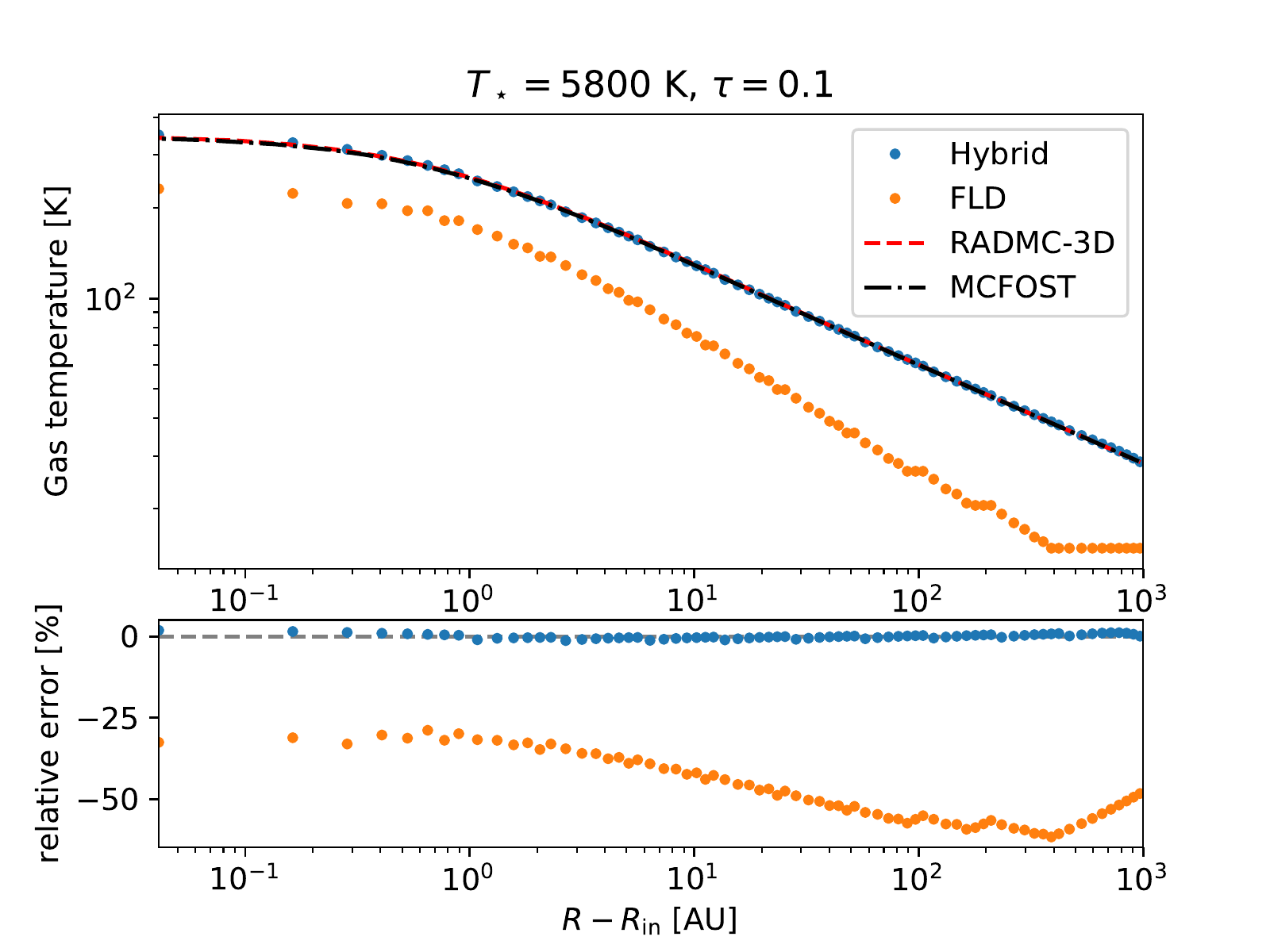}
  \includegraphics[width=9cm]{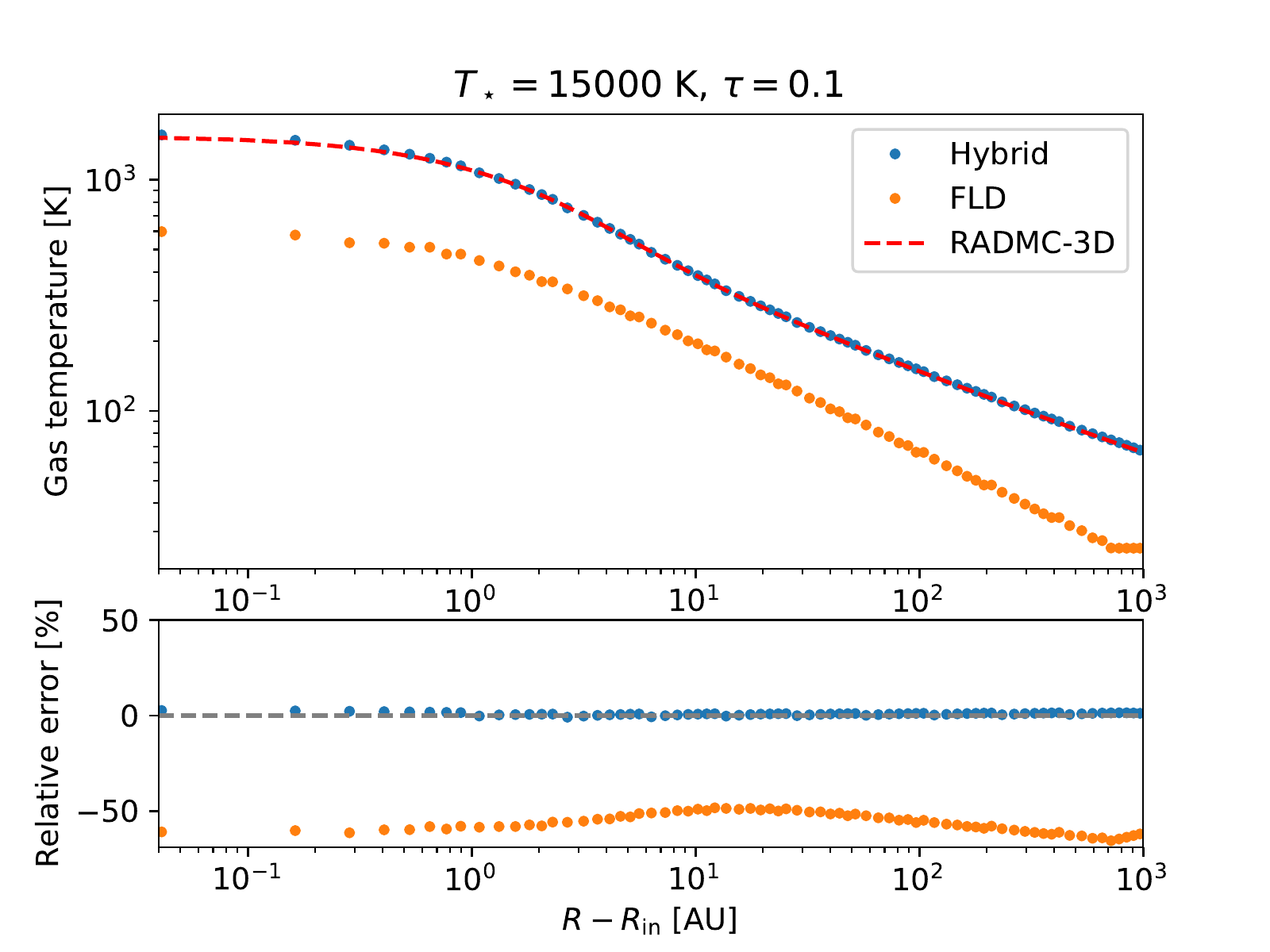}
    \caption{Radial gas temperature profiles in the mid-plane of the disk following the test of \cite{pascucci_2d_2004} for $\tau=0.1$. We compare the gas temperature computed using \textsc{MCFOST} (black dotted-line) and \textsc{RADMC-3D} (red dashed-line), the hybrid method (M1+FLD, blue dots) and the FLD method alone (orange dots) in \ramses{}. Left: central star temperature $T_\mathrm{\star,1} = 5800 \mathrm{K}$; right: $T_\mathrm{\star,2} = 15000 \mathrm{K}$.}
    \label{fig:x_T_pa_t1_m1vsfldvsmcfost}
\end{figure*}

The first test we have performed is taken from
\cite{pascucci_2d_2004} and consists of a star irradiating a static disk made of dust and gas.
We use it to probe the behavior and accuracy of our method in the optically-thin and moderately optically-thick regimes. In particular, we compare our results once the temperature structure is converged with respect to time with those obtained with Monte-Carlo RT codes such as RADMC-3D \citep{dullemond_radmc-3d:_2012} and MCFOST \citep{pinte_monte_2006}.

This is a 2D test of a static flared disk of a given analytical profile for the gas density, depending on the cylindrical radius $r$ and on the vertical height $z$.
The disk extends from $\mathrm{r_{in}} = 1 \mathrm{\, AU}$ to $\mathrm{r_{out}} = 1000 \mathrm{\, AU}$.
The density $\rho(r,z)$ in cylindrical coordinates is given as
\begin{equation}
\rho(r,z) = \rho_0 f_\mathrm{1}(r)f_\mathrm{2}(r,z),
\end{equation}

where $\rho_0$ is the density normalization and is linked to the only free-parameter, the integrated optical-depth throughout the mid-plane of the disk, $\tau_\nu = \int_\mathrm{r_{in}}^\mathrm{r_{out}} \kappa_\nu \rho(r,z=0) \, \mathrm{d}r$. The two functions $f_1$ and $f_2$ are given by
\begin{equation}
f_1(r) = \left( \frac{r}{r_\mathrm{d}} \right) ^{-1},
\end{equation}
and
\begin{equation}
f_2(r,z) = exp \left( -\frac{\pi}{4} \left( \frac{z}{h(r)} \right)^2 \right),
\end{equation}
where the flaring function is
\begin{equation}
h(r) = z_\mathrm{d} \left( \frac{r}{r_\mathrm{d}} \right) ^{1.125}.
\end{equation}

In this setup, $r_\mathrm{d} = r_\mathrm{out}/2=500 \mathrm{\, AU}$
and
$z_\mathrm{d} =r_\mathrm{out}/8=125 \mathrm{\, AU}$ are the scale-radius and the scale-height.
The star is not resolved but its luminosity is based on its physical radius and temperature.
In this test, it has a radius $R_\star = 1 \, \mathrm{R}_\odot$ and can have two possible surface temperature: $T_{\star,1} = 5800 \mathrm{\, K}$ and $T_{\star,2} = 15000 \mathrm{\, K}$.

The integrated optical depth (for extinction, as in the literature) is taken to be either $\tau=0.1$ or $\tau=100$ at $550$ nm to probe the optically-thin and moderately optically-thick regimes, respectively.
The dust-to-gas mass ratio is equal to $0.01$. Dust is made of spherical astronomical silicates of radius $0.12$ micron and density of $3.6 \mathrm{\, g\,cm^{-3}}$.
Frequency-dependent dust opacities are taken from \cite{draine_optical_1984} as in \cite{pascucci_2d_2004} and are displayed in Fig.~\ref{fig:opa_pasc}. In these setups we only take the absorption into account and do not consider scattering.
The corresponding Planck and Rosseland mean opacities used in the gray M1 and FLD modules are displayed in Fig.~\ref{fig:kp_kr_pasc}. We recall that we take the M1 absorption coefficient as the Planck mean opacity at the stellar temperature, $\kappa_\mathrm{P} \,(T_\star)$.

Boundary conditions are chosen to be a fixed temperature of $14.8 \mathrm{K}$ and a density floor of $10^{-23} \mathrm{\, g\,cm^{-3}}$. The same density floor is applied between the star and the disk edge to mimic the vacuum that \textsc{RADMC-3D} and \textsc{MCFOST} strictly apply since their respective cylindrical and spherical grids begin at $R_\mathrm{in}$. The $14.8 \mathrm{K}$ temperature is applied throughout the computational domain as initial condition and is at equilibrium with radiation.

We run the simulations with AMR levels between $5$ and $14$, which results in a finest resolution of $\mathrm{\Delta x} = 0.12~\mathrm{AU}$ where $\mathrm{\Delta x}$ is the cell width.
This makes possible to have several (${\approx}9$) cells between the star and the disk edge and the star to have a negligible size with respect to the disk thickness (${\approx}\SI{0.01}{AU}$ against ${\approx}\SI{0.04}{AU}$ for the disk height at $r_\mathrm{in}$).
Secondly, it permits to resolve several times the mean free-path at the disk inner edge: the local optical depth is $\kappa_{\mathrm{P}} \, \rho_{\mathrm{max}} \, \mathrm{\Delta x} {\approx}0.15 < 1$, where $\rho_{\mathrm{max}}$ is the density at the disk inner edge for the case $\tau=100$.
Refinement is performed on the density gradient so that the disk inner edge is at the highest refinement level.
We consider that the temperature structure is converged when the relative change between successive outputs decreases below $10^{-4}$ (see \citealp{ramsey_radiation_2015}).

\subsubsection{Temperature structure}

The \ramses{} grid is Cartesian while the grids of \textsc{MCFOST} and \textsc{RADMC-3D} are cylindrical and spherical, respectively.
Therefore, we interpolate temperature
values on their grids to compute the relative error at the location on the \ramses{} grid.
Figure~\ref{fig:x_T_pa_t1_m1vsfldvsmcfost} plots the gas temperature in the disk mid-plane against the $x$-axis for the most optically-thin case, $\tau=0.1$, once the temperature structure is converged with respect to time.
The location is given by the distance to the disk inner edge, $r-r_\mathrm{in}$.
For $T_{\star,1}$ and $T_{\star,2}$ the FLD run produces an important error throughout the disk mid-plane, up to ${\approx}62\%$ and ${\approx}65 \%$, respectively, and always underestimates the temperature. On the opposite, the hybrid method is quite accurate with a maximal error of ${\approx}2\%$ while RT codes (\textsc{MCFOST} and \textsc{RADMC-3D}) agree within $1 \%$ in this test (in accord with \citealt{pascucci_2d_2004}).
This important difference between FLD and hybrid methods comes from the regime of validity of each method: the FLD is not well-suited for optically-thin media.
In the hybrid method, since there is not much absorption because of the low optical-depth, the M1 module is mainly at work and is adapted to optically-thin media (as tested in \citealt{rosdahl_ramses-rt:_2013}), which justifies its good accuracy.

For direct irradiation, the Planck mean opacity in the FLD implementation is computed at the local temperature even though the radiation has been emitted by the star.
A direct consequence is that the stellar radiation is absorbed by the disk with an opacity coefficient computed at the disk temperature, which is much lower than the stellar temperature.
As the opacity increases with the temperature (see Fig.~\ref{fig:kp_kr_pasc}), the absorption opacity with the FLD is lower and hence the temperature is lower than that given by RT codes and by the hybrid approach.
The situation worsens from $T_\mathrm{\star,1}$ to $T_\mathrm{\star,2}$ at the disk edge and the error increases from ${\approx}30\%$ to ${\approx}60 \%$.
It also illustrates the need for a better approach for treating massive stars irradiation.

\begin{figure*}
\centering
  \includegraphics[width=9cm]{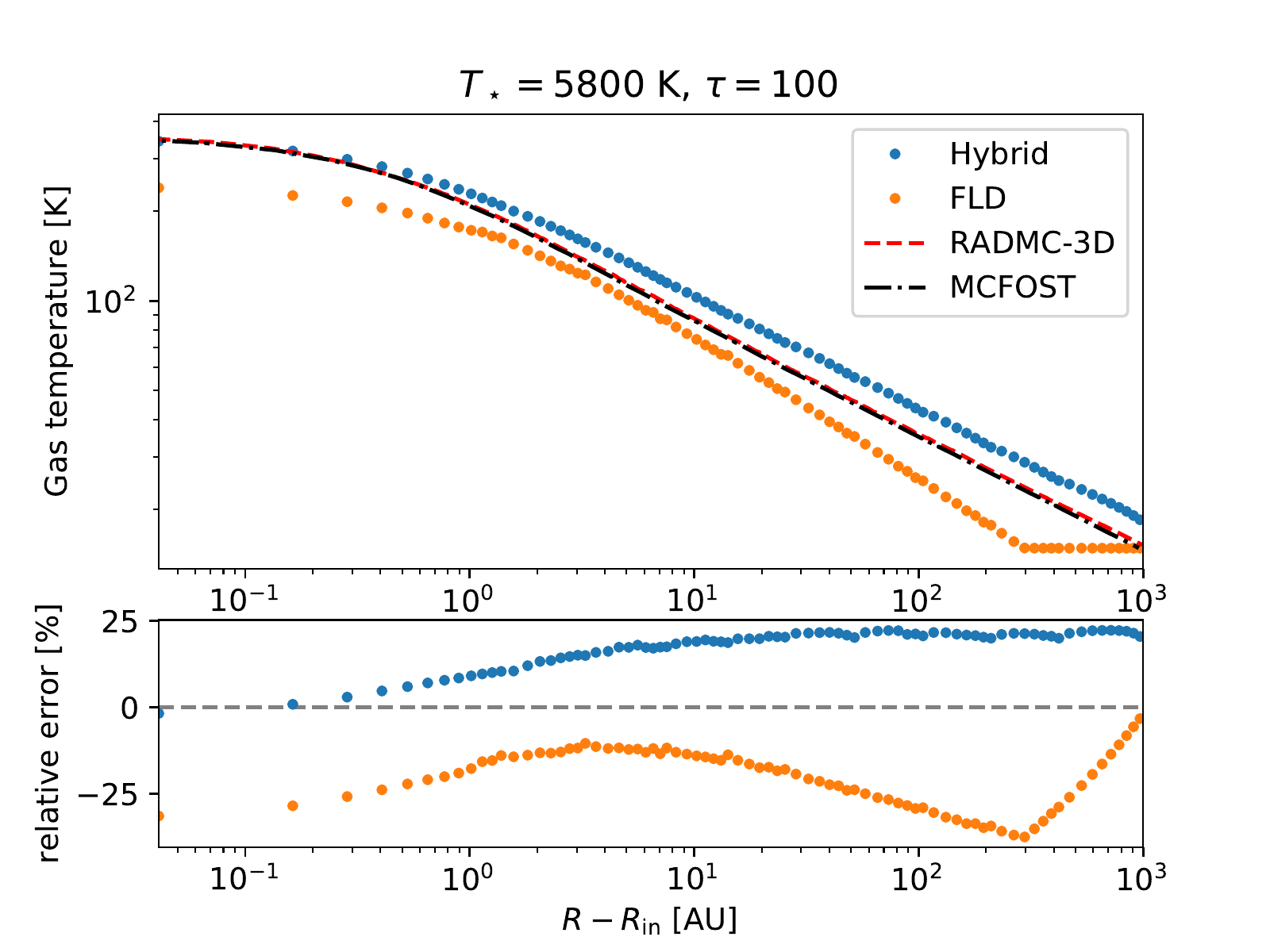}
  \includegraphics[width=9cm]{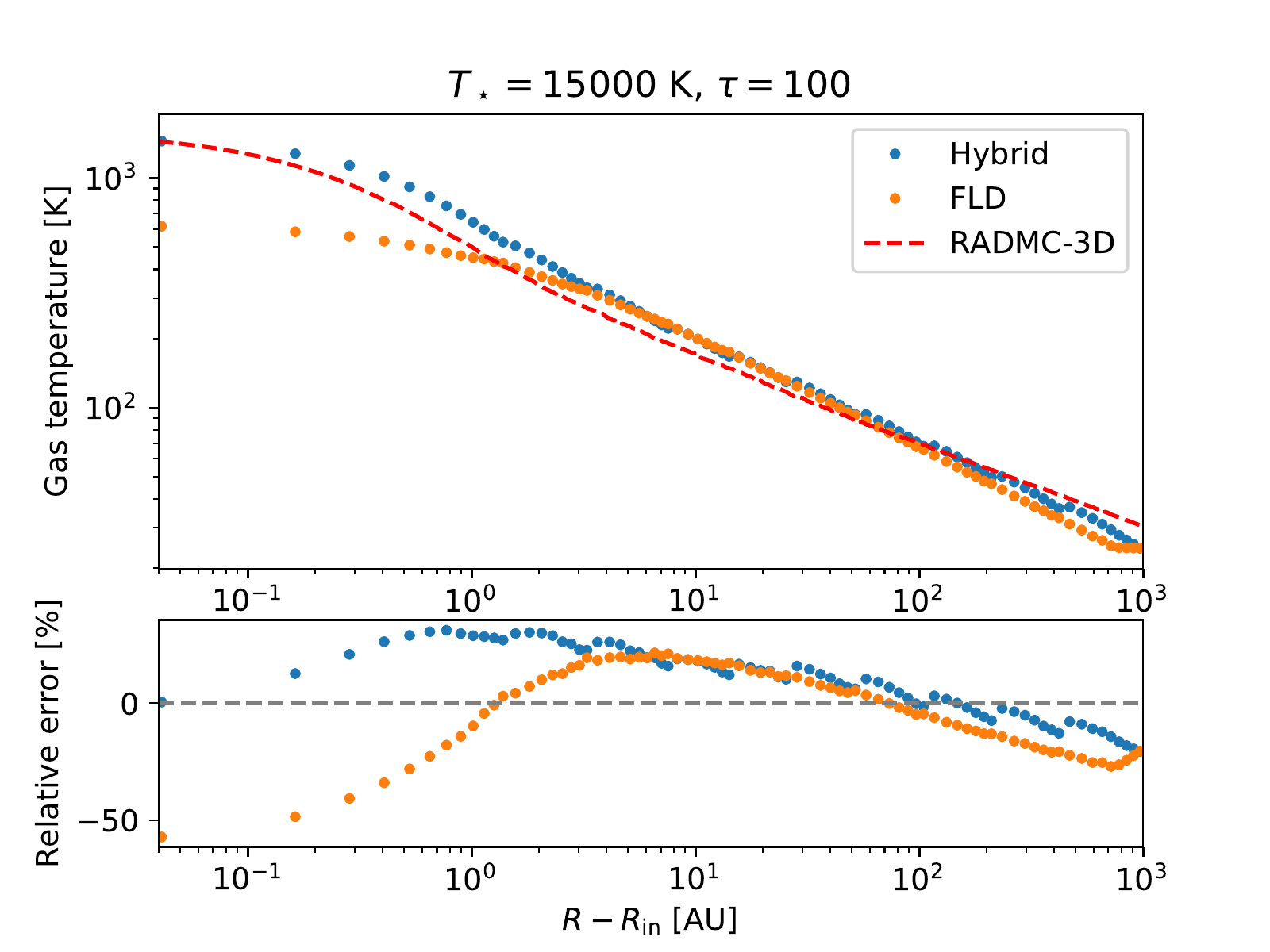}
    \caption{Same as Fig.~\ref{fig:x_T_pa_t1_m1vsfldvsmcfost}, but for $\tau=100$.}
    \label{fig:x_T_pa_t2_m1vsfldvsmcfost}
\end{figure*}

\begin{figure*}
\centering
  \includegraphics[width=9cm]{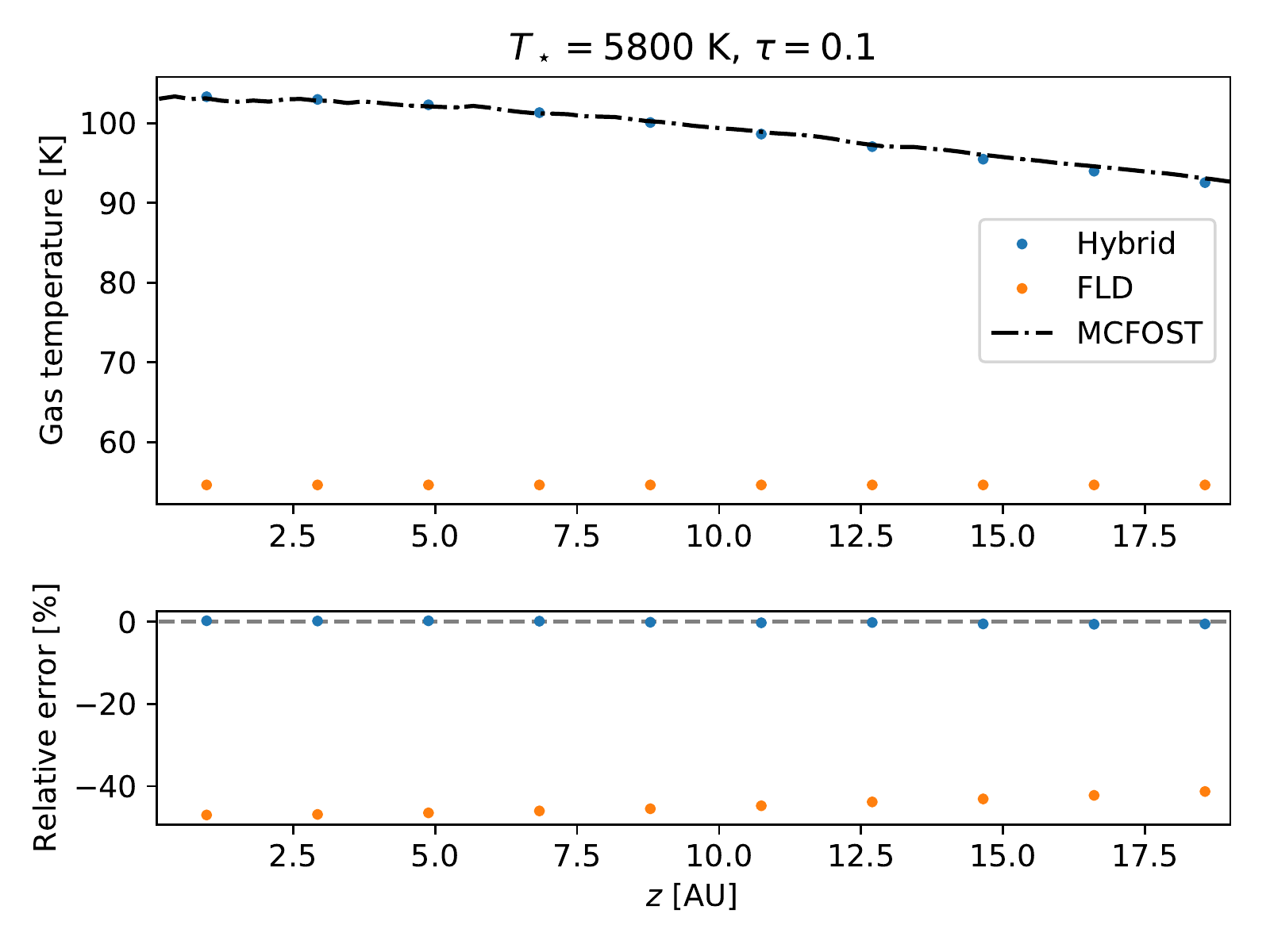}
  \includegraphics[width=9cm]{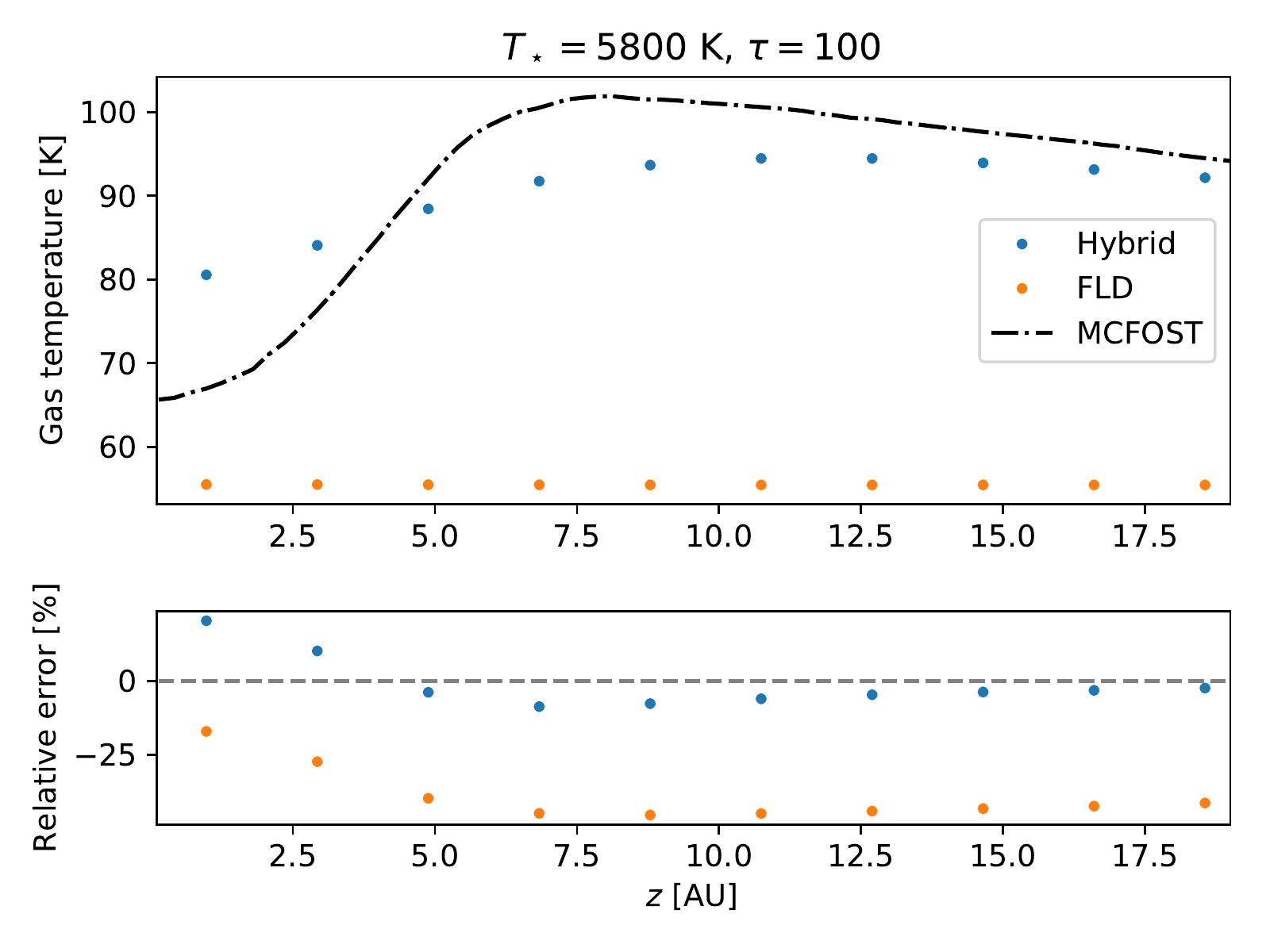}
    \caption{Vertical gas temperature profiles at a cylindrical radius of $20$ AU, following the test of \cite{pascucci_2d_2004}. We compare the gas temperature computed using \textsc{MCFOST}, the hybrid method (M1+FLD) and FLD alone in \ramses{} for $T_{\star,1}$, $\tau=0.1$ (left) and $\tau=100$ (right).}
    \label{fig:z_T_pa_m1vsfldvsmcfostnoscat}
\end{figure*}

The left panel of Fig.~\ref{fig:x_T_pa_t2_m1vsfldvsmcfost} shows the radial temperature profile in the disk mid-plane for the moderately optically-thick case, $\tau=100$, and for $T_{\star,1}$.
The error made by the hybrid method is higher than for the $\tau=0.1$ case and reaches a maximal value of ${\approx}25 \%$ whereas the FLD method alone makes a maximal error of ${\approx}36\%$.
Also, the error made by the hybrid method is quite uniform with respect to the error made by the FLD method alone.
For $T_{\star,1}$ and $T_{\star,2}$ (left and right panels of Figs.~\ref{fig:x_T_pa_t1_m1vsfldvsmcfost} and \ref{fig:x_T_pa_t2_m1vsfldvsmcfost}, respectively), the FLD method underestimates the temperature between the star and the disk edge because the medium is optically thin and because of the Planck opacity considered, as explained above. 
For $T_{\star,2}$, both methods converge toward a similar temperature at large radii.
Absorption is stronger here than in the most optically-thin case, and stronger than for $T_{\star,1}$ (see Fig.~\ref{fig:kp_kr_pasc}), so the M1 photons are quickly absorbed and the FLD module of our hybrid method is at work. Therefore a significant error is expected from the gray opacity employed in the FLD module of the hybrid method.
Indeed, the temperature varies significantly throughout the disk (between ${\approx}20$ and $350 ~\mathrm{K}$ for $T_\mathrm{\star,1}$ and between ${\approx}20$ and $1000~\mathrm{K}$ for $T_\mathrm{\star,2}$).
As a consequence, a disk cell is crossed by photons of very different frequencies and the gray approach induces errors.
Conversely, the frequency-dependence of RADMC-3D method permits one to distinguish the photons that are quickly absorbed (the most energetic ones) from those at lower energy that penetrate the disk more deeply and contribute to the disk heating at larger radii.

To examine the behavior of both methods with respect to the nonisotropy of the setup, we plotted the vertical temperature profile at a cylindrical radius of $20~\mathrm{AU}$ (Fig.~\ref{fig:z_T_pa_m1vsfldvsmcfostnoscat}).
This visualization is important for this type of tests, because an optically-thick disk produces self-shielding in the mid-plane and we expect the hybrid method to capture it better than the FLD method \citep{gonzalez_heracles:_2007}.
Here we take the temperature given by \textsc{MCFOST} rather than \textsc{RADMC-3D} because its grid is cylindrical (and not spherical) and thus errors of interpolation are avoided. 
The left panel of Fig.~\ref{fig:z_T_pa_m1vsfldvsmcfostnoscat} shows the temperature profile for the most optically-thin case, $\tau=0.1$.
No self-shielding is expected and the temperature should decrease slowly as the vertical height increases. 
Such a behavior is obtained with \textsc{MCFOST} as well as with the hybrid method. The temperature obtained with the FLD method is uniform with $z$, which is likely due to the isotropic nature of the FLD method.
The relative error is comparable to the one in the radial profile: up to ${\approx}47\%$ with the FLD method and less than $1 \%$ with the hybrid approach.

On the right panel of Fig. \ref{fig:z_T_pa_m1vsfldvsmcfostnoscat}, $\tau = 100$, and \textsc{MCFOST} gives a lower temperature in the mid-plane than for $\tau = 0.1$, as expected.
Conversely, the FLD method does not capture at all the nonisotropic nature of the radiation onto the irradiated disk: the temperature is fairly uniform.
The hybrid method reproduces partly this feature, even though the error can be as large as ${\approx}20 \%$.

We conclude that the FLD method is not capable of reproducing the temperature profile in the optically-thin and moderately optically-thick regime. The hybrid method is very accurate in the optically-thin regime (less than ${\approx}2\%$). 
In the moderately optically-thick regime, the hybrid method gives a non-negligible error (up to ${\approx}31\%$ for a $15000~$K star) in the transition between optically-thin and -thick media which shows its limitations but this is a major improvement with respect to the ${\approx}57\%$ error made with the FLD method.
In addition, the hybrid approach captures partially (${\approx}20 \%$ error) the self-shielding in the disk mid-plane while the FLD approximation does not. 

\subsubsection{Impact on the radiative acceleration}
\label{racc}

\begin{figure*}
\centering
  \includegraphics[width=18cm]{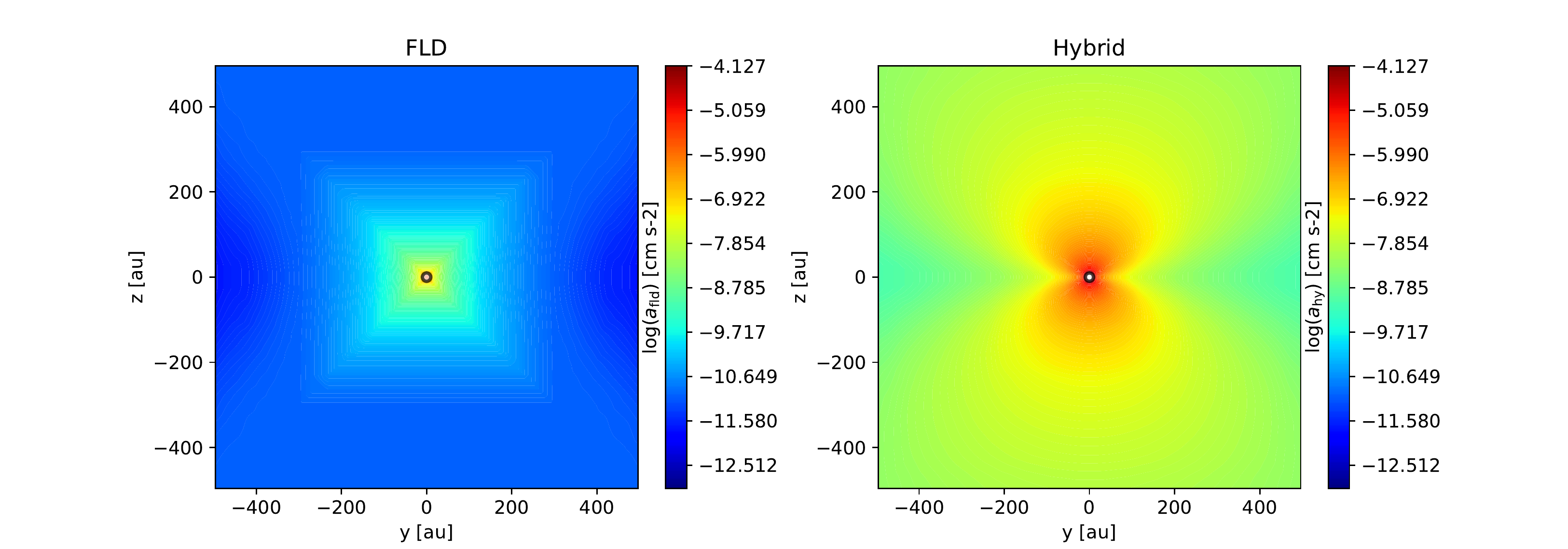}
    \caption{$1000$ AU disk edge-on slices of the radiative acceleration, following the test of \cite{pascucci_2d_2004} obtained with \ramses{} after stationarity is reached. Left: FLD run; right: hybrid run. Star and disk parameters: $\tau=100$ and $T_{\star,1}$. The hybrid radiative acceleration is about 100 times greater than the FLD one.}
    \label{fig:owen_pat2T1}
\end{figure*}

We look at the radiative acceleration maps obtained with the FLD and the hybrid methods for the moderately optically-thick case ($\tau=100$).
Figure \ref{fig:owen_pat2T1} shows the radiative acceleration perpendicularly to the disk plane as obtained after temperature convergence with the FLD method (left) and the hybrid method (right).
The left panel shows two peculiarities of the FLD solver.
First, we recall that the FLD radiative acceleration has two asymptotic values depending on the optical regime (see subsection~\ref{subs:radacc}): in the optically-thin limit it is proportional to the radiative energy and in the optically-thick limit it is equal to the radiative energy gradient divided by the density.
Further from the star, the disk structure is visible (the dark blue zones) because of the density dependence in the radiative acceleration.
Second, the aspect of the FLD acceleration closer to the star is mainly due to grid effects.

The right panel of Fig.~\ref{fig:owen_pat2T1} shows the sum of the FLD and M1 radiative accelerations in the hybrid case.
The combination of the optically-thin and -thick methods permits to capture the nonisotropy of the radiative acceleration.
The hybrid radiative acceleration is ${\sim}100$ greater than the FLD acceleration.
This result holds in the four tests: $\tau=0.1$, $\tau=100$ and  $T_{\star,1}$, $T_{\star,2}$.
It is in agreement with the study of \cite{owen_radiative_2014}.
This is mainly due to the temperature at which the M1 opacity is taken.
Stellar photons are at a frequency that is ${\sim}10$ times greater than that of photons emitted by the surrounding gas, which implies an opacity ${\sim}100$ greater (see Fig.~\ref{fig:opa_pasc}).
As shown previously, the radiation transport in the optically-thin limit is accurately treated with our hybrid approach and leads to a strong improvement for the radiative acceleration due to the direct irradiation, which is one of the main contributors expected in the dynamics of massive star formation.

\subsection{Optically-thick regime: Pinte's test}
\label{sec:pinte}

The second test is a similar but more challenging setup with a higher integrated optical depth and a sharper density profile than \cite{pascucci_2d_2004} at the disk edge, as presented in \cite{pinte_benchmark_2009}.
The disk extends from a cylindrical radius $\mathrm{r_{in}} = 0.1 \mathrm{\, AU}$ to $\mathrm{r_{out}} = 400 \mathrm{\, AU}$ and the integrated optical depth (for extinction) is $\tau_\mathrm{810 nm}=10^3$.
The flared disk density profile $\rho(r,z)$ is analytically given by
\begin{equation}
\rho(r,z) = \rho_0 \left( \frac{r}{r_\mathrm{d}} \right) ^{-2.625} exp \left( -\frac{1}{2} \left( \frac{z}{h(r)} \right)^2 \right),
\end{equation}
where the flaring function $h(r)$ is as before, $r_\mathrm{d} = r_{\mathrm{out}}/4=100 \mathrm{\, AU}$
and
$z_\mathrm{d} =r_{\mathrm{out}}/40=10 \mathrm{\, AU}$.
The star has a radius $R_\star = 2 \, \mathrm{R_\odot}$ and the stellar surface temperature is $T_{\star} = 4000 \mathrm{\, K}$.

We use the opacity table from \cite{weingartner_dust_2001} which gives the absorption opacity of dust grains with respect to the wavelength.
These opacities were calculated for spherical astronomical silicates (see \citealt{draine_optical_1984}) of size $1$ micron and density $3.6 \mathrm{\, g\,cm^{-3}}$.

\begin{figure}
    \resizebox{\hsize}{!}{\includegraphics{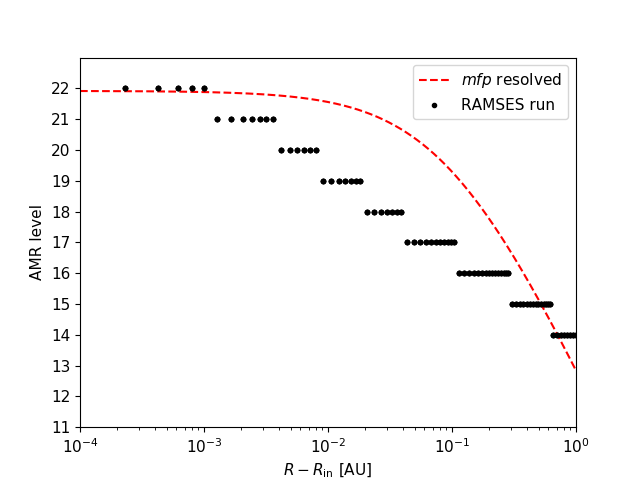}}
    \caption{AMR level needed to resolve the mean free path (mfp) of photons (red dashed-line) and effective AMR level (black dots) in the disk midplane following the test of \cite{pinte_benchmark_2009}.}
    \label{fig:x_lvl_pi_t1_T1_AMR522}
\end{figure}
\begin{figure*}
\centering
  \includegraphics[width=9cm]{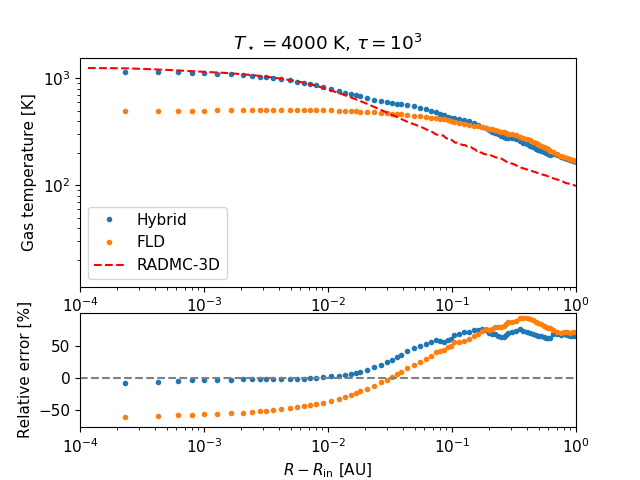}
    \includegraphics[width=9cm]{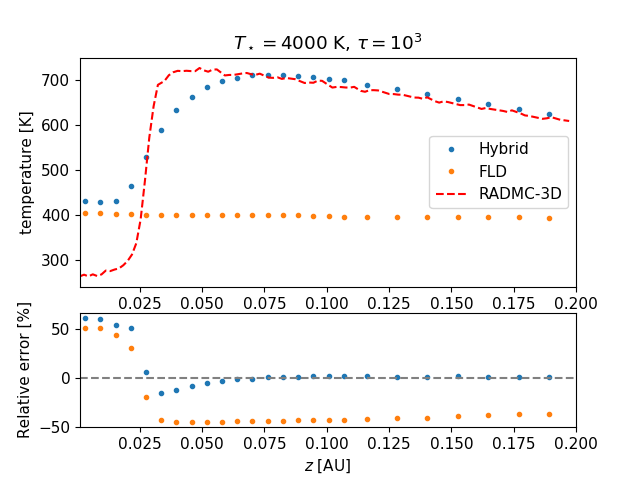}
    \caption{Left: Radial gas temperature profile in the mid-plane of the disk following the test of \cite{pinte_benchmark_2009}. Right: Vertical gas temperature profile at a cylindrical radius of \SI{0.2}{AU} in the disk. We compare the gas temperature computed using \textsc{RADMC-3D}, the hybrid method (M1+FLD) and the FLD method alone in \ramses{}. The integrated optical depth in the disk mid-plane is $\tau_\mathrm{810 nm}=10^3$ and the stellar temperature is $T_{\star}=4000~$K.}
    \label{fig:x_T_pi_t1_m1vsfldvsradmc}
\end{figure*}

The sharp increase in density at the disk inner edge makes this test particularly challenging because the local variation of optical depth must be resolved while the discretized equations involve locally constant absorption opacities.
At the same time, it is crucial to resolve the local mean free path to prevent an excess of photon absorption, which leads to overestimating the temperature.
Resolving both is even more challenging for AMR-grid codes than for cylindrical-grid codes with no material inside $\mathrm{R}_\mathrm{in}$ and a logarithmic scale.
In \ramses{} we choose to refine the grid based on a density gradient criterion so that the disk edge is at the finest resolution and the transition from optically-thin to -thick is as resolved as possible.
There is a drawback: having the greatest resolution at the disk inner edge is very computationally expensive because, first, it affects many more cells than if the refinement is operated on the central cell (as usually done because the sink particle is located there).
In addition, two adjacent cells cannot differ by more than one level of refinement and generally the number of cells at the same AMR level is much higher than two.
Therefore, it also means a higher resolution at larger radii.
For that reason, \cite{ramsey_radiation_2015} choose not to use AMR but instead, a logarithmically-scaled grid particularly adapted to this setup.
Figure~\ref{fig:x_lvl_pi_t1_T1_AMR522} shows the AMR level needed to resolve the local mean free path as a function of the radius in the disk midplane along with the AMR level set with \ramses{}.
Hence, we perform our calculations with $l_\mathrm{max} = 22$, which gives a finest cell width of \SI{1.9e-4}{AU}, so that the mean free path at the disk inner edge is resolved.
The computational cost does not make possible to extend the zone over which the mean free path is resolved, nor to compare the temperature over the entire disk radius with \textsc{RADMC-3D}.

The left panel of Fig.~\ref{fig:x_T_pi_t1_m1vsfldvsradmc} plots the radial temperature profile in the disk mid-plane obtained with \ramses{} with the FLD and hybrid methods versus \textsc{RADMC-3D}.
FLD underestimates the temperature at the disk inner edge, as in the test of \cite{pascucci_2d_2004}.
The temperature slope given by the hybrid method is in a better agreement with \textsc{RADMC-3D} than the one given by the FLD method.
For the hybrid method, the temperature at the inner edge of the disk is accurately computed (up to ${\approx}7 \%$ error) but is overestimated at larger radii where it becomes fairly constant at ${\approx}65 \%$ error.
It can be seen that the error made by the hybrid approach is not negligible as the mean free path becomes unresolved (Fig.~\ref{fig:x_lvl_pi_t1_T1_AMR522}).
The temperature profile obtained with our hybrid method is very similar to what has been obtained in comparable studies (see Fig.~8 of \citealt{ramsey_radiation_2015}).

The right panel of Fig.~\ref{fig:x_T_pi_t1_m1vsfldvsradmc} shows the vertical temperature profile.
The temperature profile shape given by the hybrid method is similar to \textsc{RADMC-3D} but with self-shielding partially captured (up to ${\approx}61 \%$ error), unlike in the FLD method.
The hybrid method then recovers the correct temperature (${\approx}2 \%$ error) at a larger disk height.

This setup highlights the need to resolve the mean free path of photons to obtain the correct temperature at the disk edge and it shows that the hybrid approach is more accurate than the FLD approximation to compute the temperature structure of an optically-thick disk.
Moreover, this setup is challenging for our hybrid method because most of the direct irradiation is absorbed in the inner parts of the disk so the rest of the disk temperature structure is mainly obtained with the FLD method.
As shown in \cite{ramsey_radiation_2015}, a frequency-dependent irradiation scheme is not more accurate in this test. 
However, a multigroup FLD method \citep{gonzalez_multigroup_2015} would improve this, as mentioned by \cite{ramsey_radiation_2015}.

\section{Collapse of an isolated massive prestellar core}
\label{sec:col}

We use the newly implemented and tested hybrid method in the context of a massive star formation and study the influence of using such a radiation transport method with respect to the previously used flux-limited diffusion approximation.

\subsection{Included physics}

Our simulations are run with the \ramses{} code \citep{teyssier_cosmological_2002} which includes a hydrodynamics solver, sink particle algorithm \citep{bleuler_towards_2014}, and radiative transfer with either the flux-limited diffusion module alone (which we call the FLD run) or coupled to \ramsesrt{} (the HY run) within our hybrid approach.
The opacities were originally used in the low-mass star formation calculations of \cite{vaytet_simulations_2013} which include frequency-dependent dust opacities ($T < 1500~\mathrm{K}$, \citealt{semenov_rosseland_2003}, \citealt{draine_scattering_2003}).
We modify the gray opacities to account for dust sublimation, because its importance for the shielding properties of massive disks has been highlighted in \cite{kuiper_role_2010}.
We model it in the same way as \cite{kuiper_circumventing_2010} (Eqs (21) and (22) therein) with a dust-to-gas ratio that decreases with temperature, and a sublimation temperature that increases with the density.
The profile of the dust-to-gas mass ratio is given by
\begin{equation}
    \frac{M_\mathrm{dust}}{M_\mathrm{gas}} (\rho, T) =\left( \frac{M_\mathrm{dust}}{M_\mathrm{gas}} \right)_0 \left( 0.5 - \frac{1}{\pi} \mathrm{arctan} \left( \frac{T-T_\mathrm{evap}(\rho)}{100} \right) \right)
\end{equation}
where $\left( \frac{M_\mathrm{dust}}{M_\mathrm{gas}} \right)_0$ is the initial dust-to-gas mass ratio,
and the evaporation temperature is given by
\begin{equation}
    T_\mathrm{evap}(\rho) = g \rho^\beta
\end{equation}
with $g = \SI{2000}{K.cm^3.g^{-1}}$, $\beta = 0.0195$ \citep{isella_shape_2005}.
At high temperature, when all dust grains are evaporated, the gas opacity is dominant and is taken equal to $\SI{0.01}{cm^2.g^{-1}}$ for comparison purposes with previous studies, such as \citealt{krumholz_formation_2009}, \citealt{kuiper_solution_2014}, \citealt{rosen_unstable_2016}, \citealt{klassen_simulating_2016}.

\subsection{Setup}

We start from initial conditions similar to \cite{rosen_unstable_2016}: a \SI{150}{\solm} spherical cloud of radius \SI{0.1}{pc} in a box of size \SI{0.4}{pc} to limit boundary effects.
The density profile is spherically-symmetric and $\rho(r) \varpropto r^{-1.5}$.
The free-fall time is then
\begin{equation}
    \mathrm{\tau_{ff}} = \sqrt{\frac{3 \pi}{32 \mathrm{G} \Bar{\rho}}} \simeq \SI{42.5}{kyr},
\end{equation}
where $\mathrm{G}$ is the gravitational constant and $\Bar{\rho}$ is the mean density computed for a uniform sphere.
The density at the border of the cloud is $100$ times the density of the ambient medium.
The cloud is in solid-body rotation around the $x$-axis with rotational to gravitational energy of ${\approx}4 \%$, typical of observed cores \citep{goodman_dense_1993}.
The initial dust-to-gas mass ratio is $\left( \frac{M_\mathrm{dust}}{M_\mathrm{gas}} \right)_0 = 0.01$.

The base resolution is level $7$ ($128^3$) and the finest resolution is $4096^3$ (\textit{i.e.}, level $12$, five levels of refinement), which gives a physical maximum resolution of $20$ AU.
In order to limit artificial fragmentation \citep{truelove_97}, we impose to have at least $12$ cells per Jeans length.
Sink particles can only form in cells refined to the highest level.
Sink creation sites are identified with the clump finder algorithm of \cite{bleuler_towards_2014}.
The clump finder algorithm marks cells whose density is above a given threshold ($\SI{3.85e-14}{g.cm^{-3}}$ in these calculations).
The marked cells are attached to their closest density peak, which form a "peak patch".
We check connectivity between the patches, then the significance of a peak patch is given by the ratio between the peak density and the maximum saddle density lying at a boundary of the peak patch.
 If the peak-to-saddle ratio is lower than a given value ($2$ here) the patch is attached to the neighbor patch of highest saddle density.
 The remaining peak patches are labeled as clumps.
Each clump must then meet two conditions to lead to a sink creation: it has to be bound and subvirial.
The region around a sink particle is also refined to the highest level.
The accretion scheme is based on a density threshold.
Consider a cell located within the accretion radius: its accreted mass by the sink is $\Delta m = \mathrm{max}(0.25 (\rho-\rho_\mathrm{sink})\times \Delta x^3, 0)$, where $\Delta x$ is the maximum resolution.
We merge sinks when they are located in the same accretion volume, while the accretion radius is set to $4 \Delta x \approx 80$~AU.
The radius and luminosity of the star mimicked by the sink are computed from the pre-main sequence evolution models of \cite{kuiper_simultaneous_2013} and depend on their time-averaged accretion rate and their mass.

\subsection{Results}
\label{sec:res}

We run one simulation with FLD only (denoted as FLD) and one with FLD+M1 (denoted as HY) until $t\simeq\SI{30}{kyr} \simeq 0.71 \tau_\mathrm{ff}$.
As the initial density profile is peaked, a sink particle is expected to form in a few kyr.
Both runs lead to the formation of several sink particles, one of which is much more massive than the other sink particles and that we refer to as the main sink or star.
A disk and radiative outflows form around the main sink.
The criteria for determining the disk and outflows are explained below.
\begin{figure*}
\centering
    \includegraphics[width=8cm]{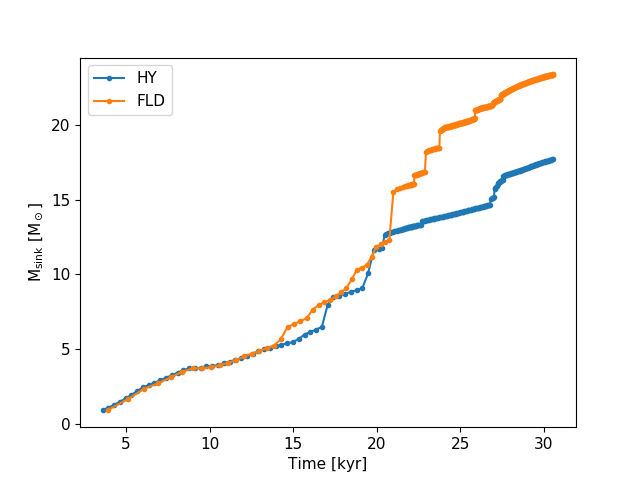}
    \includegraphics[width=8cm]{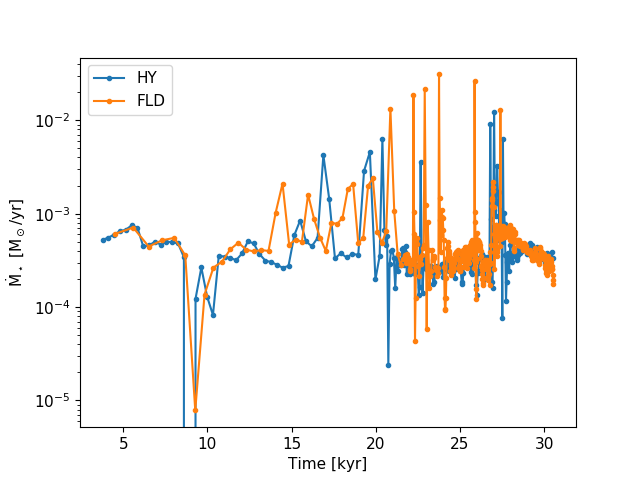}
    \includegraphics[width=8cm]{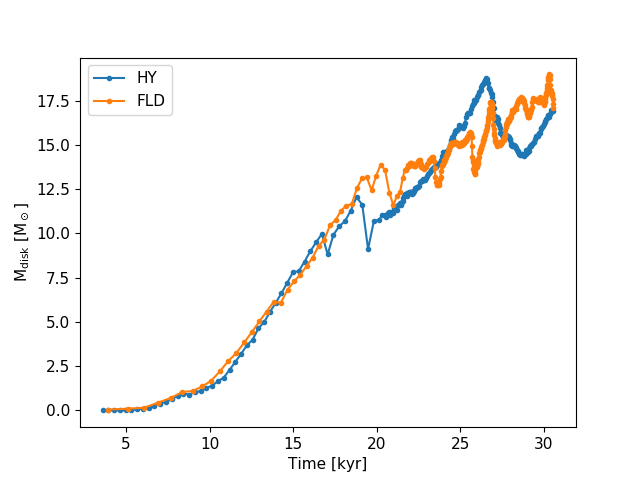}
    \includegraphics[width=8cm]{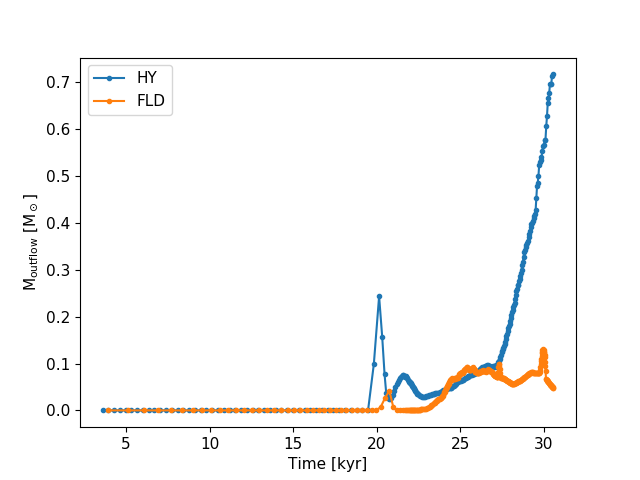}
    \caption{Time evolution of the main sink mass (top-left panel), accretion rate (top-right), disk mass (bottom-left), and outflow mass (bottom-right).}
    \label{fig:mass_evol}
\end{figure*}
We identify a disk on a cell-basis after converting Cartesian coordinates into cylindrical coordinates centered on the main sink and aligned with the rotational axis, according to the several criteria of \cite{joos_protostellar_2012}:
\begin{itemize}
    \item The disk is a rotationally-supported structure (\textit{i.e.}, not thermally supported): $\rho v_\phi^2/2 > f_\mathrm{thres} P$, where $v_\phi$ is the azimuthal velocity and $P$ is the thermal pressure. The value of $f_\mathrm{thres}=2$ is chosen, as in \cite{joos_protostellar_2012}. The use of $f_\mathrm{thres}>1$ leads to a stronger constraint on the identification of cells belonging to the disk;
    \item In order to avoid large low-density spiral arms, a gas number density threshold is set: $n>\SI{1e9}{cm^{-3}}$;
    \item The gas is not on the verge of collapsing radially: $v_\phi > f_\mathrm{thres} v_r$, where $v_r$ is the radial velocity;
    \item The vertical structure is in hydrostatic equilibrium: $v_\phi > f_\mathrm{thres} v_z$, where $v_z$ is the vertical velocity.
\end{itemize}
We define outflows as gas flowing away from the central star at a velocity greater than the escape velocity, which corresponds to $v_r > v_\mathrm{esc}=  \sqrt{2 G \mathrm{M_\star}/r}$, where $r$ is the distance between the sink and the cell and $\mathrm{M_\star}$ is the sink mass.

The time evolution of the main sink, disk and outflow masses, along with the accretion rate, are displayed in Fig.~\ref{fig:mass_evol}.
First, the sink masses are almost equal in both runs before $t \simeq \SI{14}{kyr}$ and $\mathrm{M_\star} = \SI{5}{\solm}$.
Even though their evolution differs between $\SI{14}{kyr}$ and $\SI{20}{kyr}$, their values remain similar and the divergence appears only at $t \simeq \SI{20}{kyr}$, when $\mathrm{M_\star} = \SI{12}{\solm}$ (in the HY run).
At that point, the radiative cavities appear in the HY run (see the bottom-right panel of Fig.~\ref{fig:mass_evol}).
They appear in the FLD run after the massive star has reached $\SI{16}{\solm}$.
From this time on, the sink mass increases more slowly in the HY run, this can be seen on the accretion rate (top-right panel of Fig.~\ref{fig:mass_evol}).
The final stellar mass is $\mathrm{M_\star}=\SI{23.3}{\solm}$ in the FLD run and $\SI{17.6}{\solm}$ in the HY run.

As it is shown on the top-right panel of Fig.~\ref{fig:mass_evol}, the stars experience bursts of accretion separated by ${\sim}100$ yrs to a few kyr.
These bursts are due to a low-mass companion (sink particle) being accreted by the most massive star.
In each run, the main sink experiences accretion rates of $\mathrm{M_\odot} {\sim}10^{-4}-10^{-2} \solm.\mathrm{yr^{-1}}$, which is consistent with previous numerical studies \citep{klassen_simulating_2016} and observations (review by \citealt{motte_high-mass_2018} and references therein).
In total, eight companions are formed and accreted in each run. These accretion events contribute to a total of \SI{6.7}{\solm} in the FLD run and \SI{3.9}{\solm} in the HY run, hence about $28\%$ and $22\%$ of the final primary star mass, respectively. All sinks appeared after the primary mass was greater than \SI{10}{\solm} and were accreted in less than $3$ kyr (except one, formed at large radius and which does not fall directly onto the primary sink). In each run, the most massive secondary is \SI{\sim~2}{\solm} and gathers most of its mass when orbiting close to the primary, in the disk densest regions. Our merging criterion can lead to overestimating the mass of the primary star and affect the system multiplicity. Assessing the impact of the merging criterion would require a dedicaded study, which is beyond the scope of this paper.

\subsubsection{Disk properties}

As shown in the bottom-left panel of Fig.~\ref{fig:mass_evol}, the disks obtained are massive (${\approx}\SI{17}{\solm}$ at the end of the simulation) and similar in mass in both runs.
Observational constraints on the disk mass remain sparse \citep{motte_high-mass_2018} but the disk mass we obtain is consistent with the previous numerical work of \cite{klassen_simulating_2016}.

\begin{figure}
    \resizebox{\hsize}{!}{\includegraphics{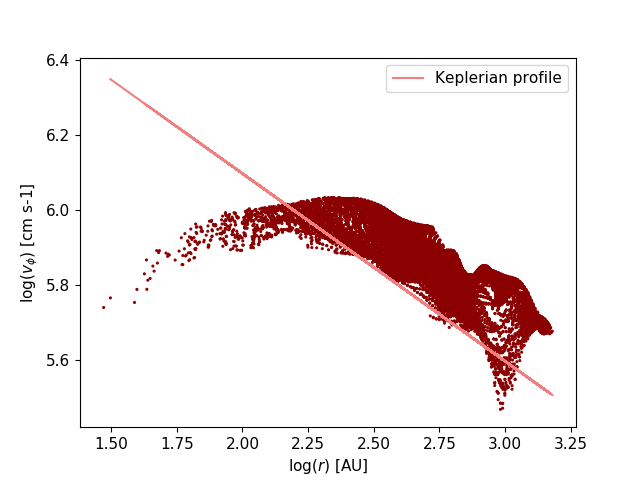}}
    \caption{Radial rotational velocity profile in the disk cells for the HY run at $t=\SI{30}{kyr}$ and Keplerian profile computed with the main stellar mass. The slope of the velocity profile is consistent with Keplerian rotation.}
    \label{fig:vphi}
\end{figure}

\begin{figure*}
\centering
    \includegraphics[width=8.7cm]{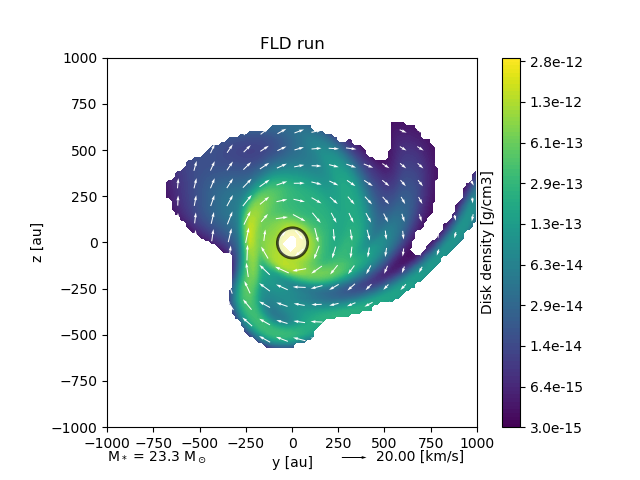}
    \includegraphics[width=8cm]{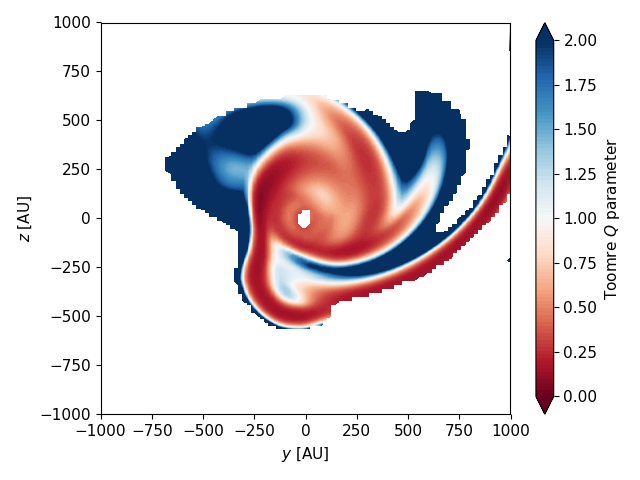}
    \includegraphics[width=8.7cm]{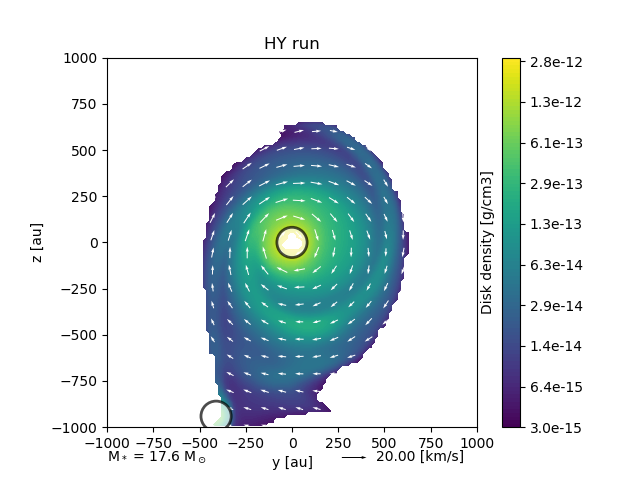}
    \includegraphics[width=8cm]{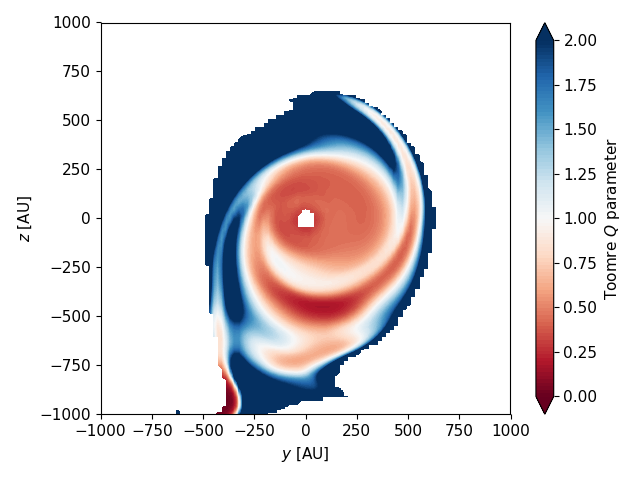}
    \caption{Density slices of the disk selection (left panels) and Toomre $Q$ parameter (right panels) in a $\mathrm{(2000 \, AU)^2}$ region centered on the location of the most massive sink particle (left panels), in the FLD run (top) and the HY run (bottom), at $t=\SI{30}{kyr}$.}
    \label{fig:Q}
\end{figure*}
We investigate the disk stability by computing the Toomre parameter $Q$ defined by
\begin{equation}
Q = \frac{c_\mathrm{s} \kappa}{\pi G \Sigma}
\end{equation}
where $c_s$ is the sound speed, $\kappa$ is the epicyclic frequency and is equal to the rotation frequency for a Keplerian disk and $\Sigma$ is the column density.
We recall that the Toomre parameter computes the ratio of the thermal support and differential rotation support over gravitational fragmentation and that the disk is locally unstable if $Q<1$.
The gas in our simulation is initially in solid-body rotation but the disks formed indeed exhibit rotation curves consistent with Keplerian rotation, as shown in Fig.~\ref{fig:vphi}.
As a result, the epicyclic frequency $\kappa$ is equal to $\Omega$, the Keplerian rotational frequency.

To calculate $Q$, we have taken the column density integrated over the $x$-axis (perpendicular to the disk).
Moreover, the selection given by the criteria presented above gives a disk with a vertical structure.
Therefore, we evaluate the Toomre $Q$ parameter in the disk selection, then we average $Q$ over the disk height. 
For completeness, we have also computed $Q$ with $c_\mathrm{s}$ and $\kappa$ evaluated in the disk midplane and have obtained very similar results.

We also take the radiation into account as an extra-support against fragmentation because the radiative pressure contributes to the sound speed (\citealt{mihalas_foundations_1984}, eq.~(101.22) therein)
\begin{equation}
c_\mathrm{s}^2 = \Gamma_1 \frac{P+P_\mathrm{r}}{\rho}
\end{equation}
where $P$ is the gas pressure, $P_\mathrm{r}$ is the radiative pressure; $\Gamma_1 = 5/3$ for a non-radiating fluid ($P_\mathrm{r} =0$, pure hydrodynamical case), and $\Gamma_1 \simeq 1.43$ if $P_\mathrm{g}=P_\mathrm{r}$).
Therefore, we argue that even for a disk in a strong radiation field and gas-radiation coupling ($P_\mathrm{r} {\sim}P_\mathrm{g}$), $Q$ only increases by a factor of $\simeq 1.3$ as compared to the pure hydrodynamical case.
Figure~\ref{fig:Q} displays the local Toomre $Q$ value taking the radiative support into account but the values of $Q$ without the radiative support lead to the same conclusions.

As shown in Fig.~\ref{fig:Q}, the disks obtained in both runs are Toomre unstable close to the massive star and in the spiral arms.
This is consistent with the regular creation of sink particles in those spiral arms.
Eight low-mass short-lived companions are generated in both runs.
Even though the appearance of sink particles is quite resolution-dependent, we limit it with our refinement criterion based on the Jeans length.

\begin{figure*}
\centering
    \includegraphics[width=8cm]{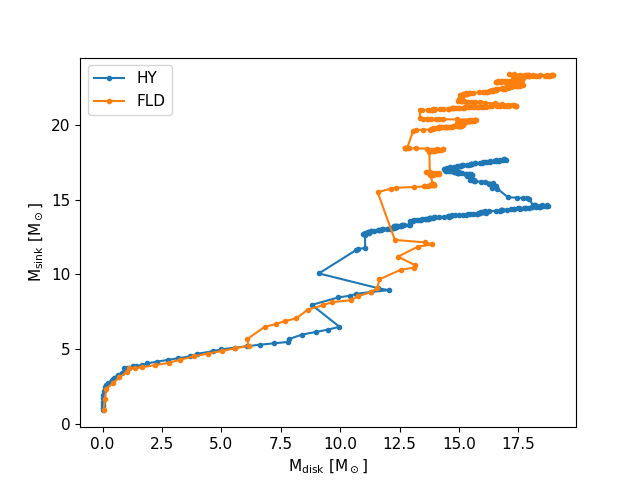}
    \includegraphics[width=8cm]{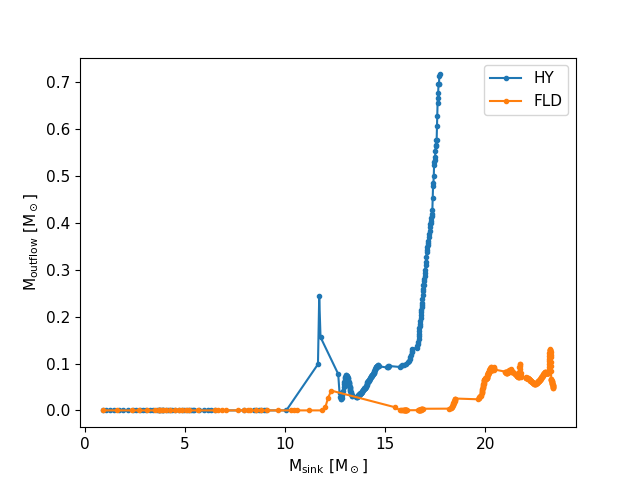}
    \caption{Primary star mass versus disk mass (left) and outflow mass versus star mass (right), for both FLD and HY runs.}
    \label{fig:masses_rel}
\end{figure*}

The left panel of Fig.~\ref{fig:masses_rel} shows the main star mass against the disk mass.
Both generally increase with time but the disk also undergoes losses of mass as it feeds the main sink particle.
Indeed, the main accretion mode in our simulation is disk accretion, although the accretion bursts are due to the accretion of sink particles recently created in the Toomre unstable spiral arms of the disk.
The accretion in our simulation is more stable than what is obtained in the work of \cite{klassen_simulating_2016}, where the global disk instability leads to an increase of $\simeq \SI{10}{\solm}$ in a few kyr in their $\SI{100}{\solm}$ run.

\begin{figure*}
\centering
    \includegraphics[width=9cm]{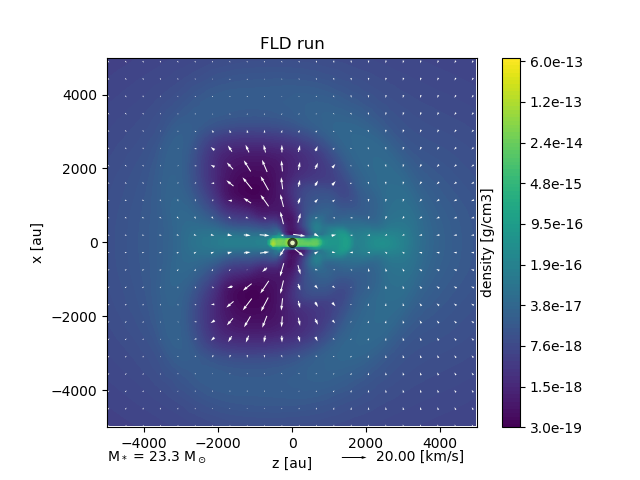}
    \includegraphics[width=9cm]{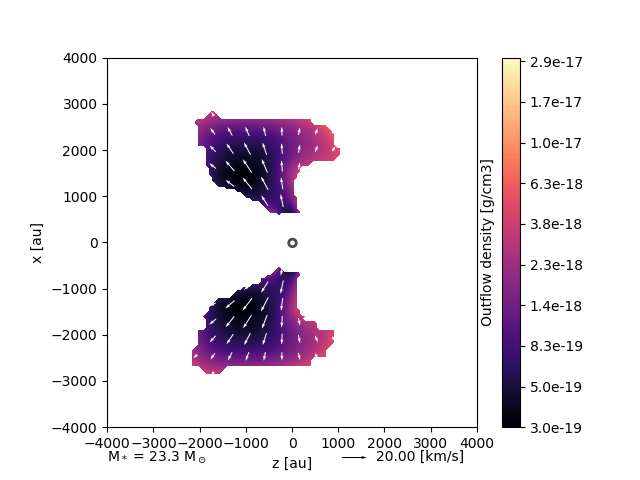}
    \includegraphics[width=9cm]{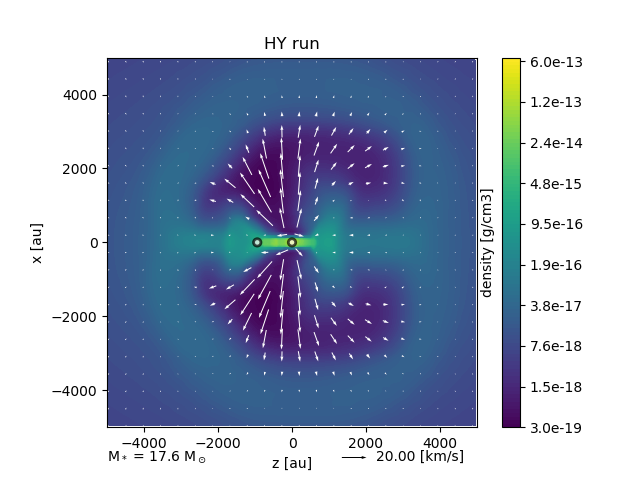}
    \includegraphics[width=9cm]{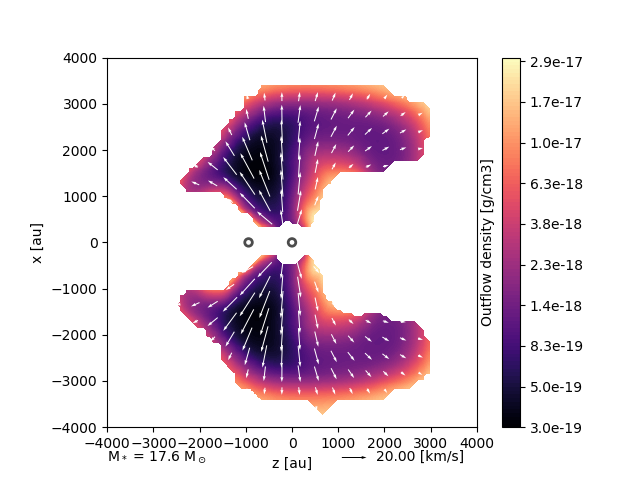}
    \caption{Left panels: density slices perpendicular to the disk in a $\mathrm{(10000 \, AU)^2}$ region. Right panels: density in outflow selections in a $\mathrm{(8000 \, AU)^2}$ region. Top panels: FLD run; bottom panels: HY run. $t=\SI{30}{kyr}$. Figures are centered on the location of the most massive sink particle.}
    \label{fig:outfl}
\end{figure*}

\begin{figure*}
\centering
    \includegraphics[width=9cm]{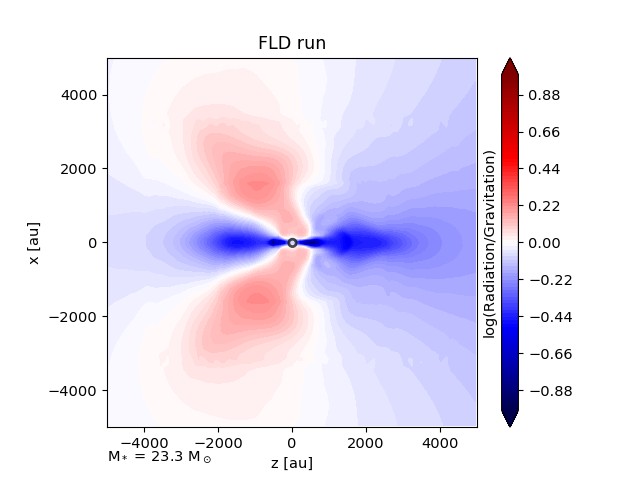}
    \includegraphics[width=9cm]{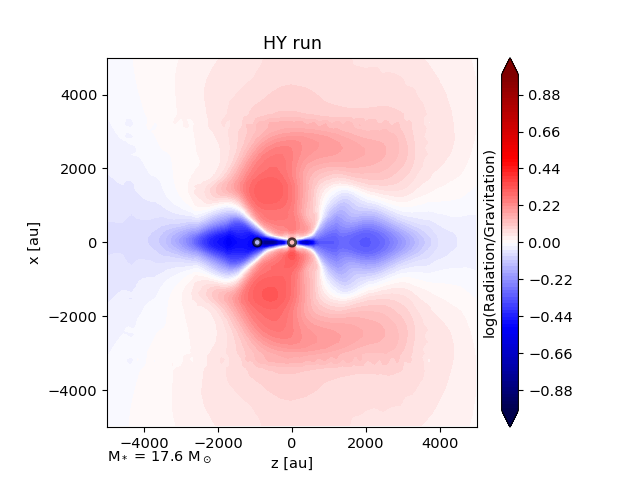}
    \caption{Radiative force to gravitational force normalized ratio in the FLD run (left panel) and HY run (right panel) in a $\mathrm{(10000 \, AU)^2}$ region perpendicular to the disk. $t=\SI{30}{kyr}$. Figures are centered on the location of the most massive sink particle. Regions of outflows (see Fig.~\ref{fig:outfl}) are dominated by the radiative force.}
    \label{fig:frad_collapse}
\end{figure*}

\subsubsection{Radiative cavities - outflows}

As mentioned in Sec.~\ref{sec:res}: we define outflows as gas flowing away from the central star at a velocity greater than the escape velocity.
In the FLD run, radiative cavities appear at $t \simeq 22$ kyr (bottom-right panel of Fig.~\ref{fig:mass_evol}).
They develop earlier and at lower stellar mass ($M_\star = \SI{12}{\solm}$, $t \simeq 20$ kyr) in the HY run than in the FLD run ($M_\star= \SI{16}{\solm}$, right panel of Fig.~\ref{fig:masses_rel}).
In both runs, the cavities grow symmetrically with respect to the disk plane until they reach an extent of $\simeq \SI{2000}{AU}$ in the FLD run and $\simeq \SI{3000}{AU}$ in the HY run at $t\simeq 30$ kyr (see Fig.~\ref{fig:outfl}).
The right panels of Fig.~\ref{fig:outfl} display a slice of the density within the outflow selection of cells.
The gas velocity is also higher in the HY run, ${\approx}\SI{25}{km/s}$, against ${\approx}\SI{15}{km/s}$ in the FLD run.
As displayed in Fig.~\ref{fig:frad_collapse}, gas is pushed away by the radiative force, which locally exceeds gravity.
It illustrates the flashlight effect: the radiative force dominates in the poles while the gravity, and hence the accretion, dominates in the disk plane.
The consequence of the stronger radiative force is that the outflows in the HY run are able to transport higher density gas than in the FLD run (see right panels of Fig.~\ref{fig:outfl}), mainly because it spans a wider angle, particularly in the vicinity of the star (see Fig.~\ref{fig:frad_collapse}).
Indeed, the outflows displayed in Fig.~\ref{fig:outfl} have masses $M_{\mathrm{o,HY}} \simeq \SI{0.6}{\solm} > M_{\mathrm{o,FLD}} \simeq \SI{0.06}{\solm}$, as displayed on the bottom-right panel of Fig.~\ref{fig:mass_evol}.
During almost all the simulations the outflows in the HY run are more massive than in the FLD run. In addition, the temporal evolution at $t\simeq \SI{30}{kyr}$ seems to show that this mass is still going to increase in the HY run but not in the FLD run.
We finally note that the peak in the outflow mass at $t\simeq \SI{20}{kyr}$ is due to the launching of the outflows in a high-density medium close to the star, which therefore gives a higher outflow mass: the low-density cavity has not formed yet.

\subsubsection{Rayleigh-Taylor instabilities}

No Rayleigh-Taylor instabilities appear in the aforementioned runs (FLD and HY).
They have been shown to contribute significantly to the star-disk system evolution and to the final mass of the star in several studies (\citealp{krumholz_formation_2009}, \citealp{rosen_unstable_2016}, \citealp{rosen_massive_2019}).
Discussions about the presence of these instabilities in massive star formation simulations lean on arguments of numerical resolution, since the smaller-scale modes are the most unstable \citep{jacquet_radiative_2011}.
Here we try to tackle this problem by cranking up the resolution to see if we get any. 

We conduct a run whose spatial resolution permits one to resolve the seeds of radiative Rayleigh-Taylor instabilities.
We call this run HY-RTi, based on the HY run restarted at the time when radiative cavities appear.
We rely on the AMR framework to resolve the radiative cavities interfaces with a refinement strategy based on the gradient of the stellar radiation
\begin{equation}
    \frac{\nabla E_\mathrm{M1} \Delta x}{E_\mathrm{M1}} < 10\%,
\end{equation}
where $E_\mathrm{M1}$ is the M1 module radiative energy and $\Delta x$ the cell width.
This means that if the radiative energy of the M1 module changes by more than $10\%$ between two adjacent cells, these cells are flagged for refinement.
We add a second and a third conditions to flag a cell: $E_\mathrm{M1} > E_\mathrm{thres} = \SI{3e-12}{erg.cm^{-3}}$, and $\norm{x} > 1500$~AU (over and under the sink-disk plane). 
The last condition is applied once the cavities are developed beyond this height.
The second and third criteria are necessary to not over-refining other regions than the interfaces of the cavities, which would explode the cost of the simulation.

\begin{figure*}
\centering
    \includegraphics[width=9cm]{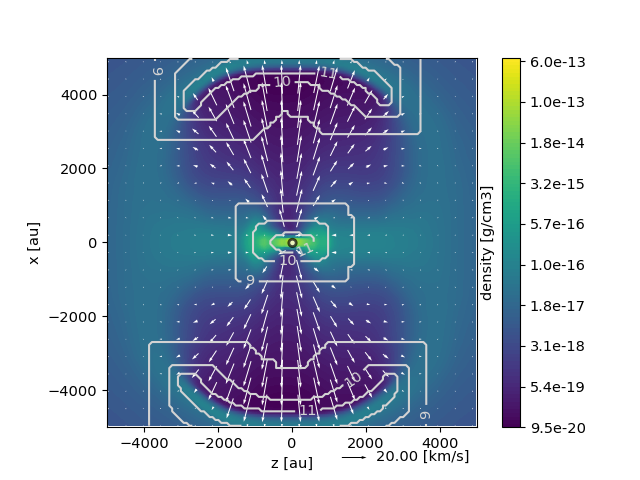}
    \includegraphics[width=9cm]{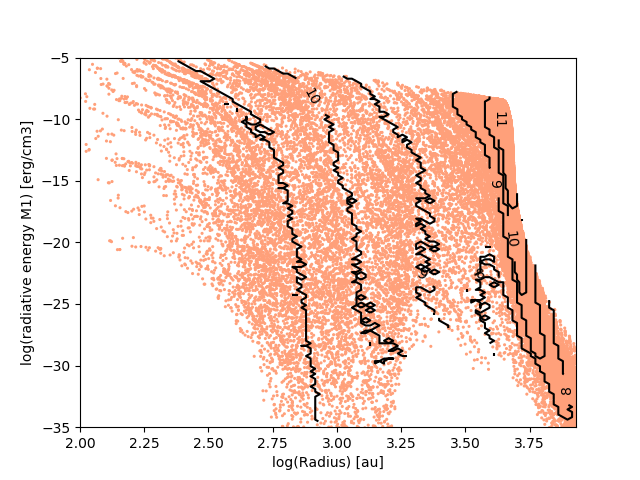}
    \caption{HY-RTi run at $t{\sim}0.7 \tau_\mathrm{ff}$. Left panel: density slice perpendicular to the disk in a $\mathrm{(10000 \, AU)^2}$ region. Right panel: scatter plot of the M1 radiative energy against the radius. Contours show the AMR level. The cavity edges are zones of primary absorption for the stellar radiation and are resolved to the highest level ($12$).}
    \label{fig:hyem1}
\end{figure*}

The left panel of Fig.~\ref{fig:hyem1} shows the radiative cavities in the HY-RTi run with AMR level contours overplotted and the right panel of Fig.~\ref{fig:hyem1} displays the M1 radiative energy with respect to the radius.
The zone within the contours "$11$" is at the AMR level $12$, which is the highest level.
Contours show that the radiative energy drop around $R = 5000$ AU is resolved to the AMR level $12$, which is the highest level.
As a result, the finest resolution is put on the cavity interfaces.
Despite this refinement strategy, no radiative Rayleigh-Taylor instability has developed in any of our simulations.
We explain below why this result is not numerical but physical.

We compare the typical advection time of the flow $\tau_\mathrm{adv}$ and the growth time of the instability $\tau_\mathrm{instab}$;
the condition for the instability to develop is $\tau_\mathrm{adv} > 3 \tau_\mathrm{instab}$ \citep{foglizzo_06}.
First, the flow in the bubble has supersonic speeds and forms a shock of thickness $H {\simeq} \SI{300}{AU}$ (measured on the density profile) when it encounters the accretion flow. In the shock frame (whose velocity is $\SI{2}{km.s^{-1}}$), we measure a gas velocity of $\SI{10}{km.s^{-1}}$. Hence, the advection timescale of the gas in the shock is $\tau_\mathrm{adv} {\simeq}  0.1$ kyr.

We now compute the growth rate of the Rayleigh-Taylor instability for the shortest perturbations we can capture (of spatial scale $\lambda = \Delta x_\mathrm{min} = \SI{20}{AU}$, which is the fastest growing mode) using the equation (80) from \cite{jacquet_radiative_2011}.
We obtain a growth rate of $\omega {\simeq}  \SI{4.5}{kyr^{-1}}$, hence a growth timescale of $\tau_\mathrm{instab} {\simeq}  0.2$ kyr.
This is longer than the advection timescale $\tau_\mathrm{adv}$, so the gas is advected before the instability develops.
Furthermore, the calculation in \cite{jacquet_radiative_2011} is based on the adiabatic approximation which is valid when the cavity edge temperature is taken to be equal to the dust sublimation temperature (${\sim} \SI{1100}{K}$).
Numerically, we get a temperature of a few ${\sim} \SI{100}{K}$ at the cavity edge, thus the adiabatic approximation breaks down in our simulation.
Physically, the cavity edge is mainly heated by stellar radiation, which is geometrically diluted in the optically-thin cavity.
Therefore, it can be shown that the cavity edge should have a temperature of a few $\SI{100}{K}$ at a distance of ${\sim} 3000$ AU from a ${\sim} 10^5 \, \mathrm{L_\odot}$ source.
Moreover, the cavity interior is optically-thin and thus is not adiabatic, as mentioned in \cite{jacquet_radiative_2011}. If compressed, the gas radiates away its energy instead of heating as an optically-thick gas (adiabatic) would.

For these reasons, we go one step further and relax the adiabatic approximation in the cavity interior. 
Hence, the entropy within the cavity cannot account for radiation. We compute the total entropy (gas plus radiation) as a function of the coupling between gas and radiation via the local optical depth $\tau$
\begin{equation}
s_\mathrm{tot} = \frac{k_\mathrm{B}}{m (\gamma -1) } \ln \left(P_\mathrm{g} \, \rho^{-\gamma} \right) + \min(\tau,1) \frac{4 P_\mathrm{r}}{\rho T},
\end{equation}
where $k_\mathrm{B}$ is Boltzmann's constant, $m$ is the molecular hydrogen mass, $P_\mathrm{g}$ is the gas pressure, and $P_\mathrm{r}$ is the radiation pressure.
The maximum growth rate is given by the Brunt-V\"ais\"ala (or buoyancy) frequency, which is the oscillation frequency of a fluid particle in a stratified medium
\begin{equation}
\omega = \sqrt{\frac{\gamma-1}{\gamma} g_\mathrm{eff} \nabla S},
\end{equation}
where $S$ is the total entropy $s_\mathrm{tot}$ normalized by $k_\mathrm{B}/m$ and
$g_\mathrm{eff}$ is the effective gravity $g_\mathrm{eff} = g - \kappa F /\mathrm{c}$, with $\kappa F /\mathrm{c}$ the radiative acceleration. 
We compute $\omega$ in our simulation at the bubble edge and get $\omega {\lesssim} \SI{10}{kyr^{-1}}$, which gives $\tau_\mathrm{instab} {\gtrsim} \SI{0.1}{kyr}$ ${\simeq} \tau_\mathrm{adv}$.
Therefore, no Rayleigh-Taylor instability should develop in our simulation.

\section{Conclusions and discussion}
\label{sec:ccl}

We have implemented a new hybrid radiative transfer method in the AMR code \ramses{} based on the flux-limited diffusion (FLD) module \citep{commercon_radiation_2011} and M1 module in \ramsesrt{} \citep{rosdahl_ramses-rt:_2013}, in order to treat accurately both the stellar irradiation and the diffuse component around a massive protostar.
Our hybrid approach takes advantage of the M1 module fully tested in the optically-thin regime and of the FLD module in the optically-thick limit.
Moreover, in contrast to the consideration of local photons inherent to the FLD approach, our method keeps the frequency information of the stellar photons propagated with \ramsesrt{}, which leads to an improvement in the treatment of coupling with the dust-gas mixture via the temperature and the radiative force.

We tested this improvement in pure radiative transfer tests of a star irradiating a static disk structure (\citealt{pascucci_2d_2004}, \citealt{pinte_benchmark_2009}).
Our results show that the hybrid method is very accurate is the optically-thin regime (${\approx} 2\%$ maximal error, ${\approx} 62 \%$ with the FLD method alone), and more accurate than the FLD method in the optically-thick regime (${\approx} 25\%$ instead of ${\approx}36 \%$ with the FLD method).
It is also capable of capturing partially the self-shielding in the disk mid-plane, because it is shielded from stellar radiation by the disk inner region.
Therefore, the hybrid method is suited to determine accurately the gas temperature structure in different regimes of optical thickness.
In addition, because stellar photons are treated apart from the photons emitted by the dusty disk, the associated radiative force is computed more consistently with the hybrid method and its value is about ${\sim}100$ times greater than with pure FLD.
This shows the need for such an hybrid method, beyond the diffusion approaches.

After testing our hybrid approach in pure radiative transfer cases, we have applied it to a radiation-hydrodynamical problem: the collapse of a massive prestellar core.
Both runs lead to the formation of a massive star.
The multi-dimensionality of our simulations leads to accretion via the flashlight effect: accretion occurs through a disk while radiation escapes via the poles \citep{yorke_formation_2002}.
Low-mass sink particles are created in the Toomre unstable spiral arms of the disk, they move together with the fluid and are rapidly accreted onto the central massive star. 
At the end of the simulation ($t \simeq \SI{30}{kyr} \simeq 0.71 \tau_\mathrm{ff}$), the star mass is $\SI{23.3}{\solm}$ in the FLD run and $\SI{17.6}{\solm}$ in the HY run,  showing no signs of decrease in the accretion.
This difference is explained by the direct radiative pressure onto the disk, which lowers the accretion rate.
When the star reaches $\SI{12}{\solm}$ (HY run) or $\SI{16}{\solm}$ (FLD run), radiative polar cavities develop because of the stellar radiative pressure.
Radiative outflows in the HY run are ${\sim}50\%$ more extended than in the FLD run.

Our method also contains a few assumptions and limitations we shall discuss.
As shown above, the hybrid method is accurate within ${\approx}25-65\%$ in the optically-thick limit. In the prestellar core collapse problem, the accretion disk around the protostellar source is very optically-thick (${\gtrsim} 10^3$) and the photon mean free path is barely resolved in AMR codes with current computational facilities. The disk midplane temperature is therefore affected by this error: it is generally overestimated, which increases the Jeans length and therefore stability.
Yet, the hybrid method clearly performs better than the pure FLD approach in predicting the mid-plane temperature.
Also, the temperature in the optically-thin cavities is computed more accurately with the hybrid method.

In this study, we have focused on a gray approach for both the direct and the diffuse radiations.
However, the gray radiative transfer is not inherent to our model and multifrequency methods have been implemented in \ramsesrt~\citep{rosdahl_ramses-rt:_2013} and in the FLD module \citep{gonzalez_multigroup_2015}.
We have chosen to work with the gray methods to save computational time and memory.
The multifrequency version of our hybrid method is beyond the scope of this paper.
We have also focused on the irradiation by one source with the M1 method.
However, the several sinks produced in these simulations were rapidly accreted by the primary one, so this does not change our conclusions. 
Moreover, the irradiation can be generalized to several sources via one photon group per frequency band or per source, and we leave this to further work.
Finally, ionization processes were not taken into account here.
They can be relevant for massive star formation, but the opening of the ionized cavities and the disk photo-evaporation has been shown to occur toward the end of our simulation \citep{kuiper_first_2018}. These physics will be included in further works.

\ramses{} includes magneto-hydrodynamics \citep{fromang_high_2006}, and the interplay between the radiation (modeled with FLD) and the magnetic field has been proven:  magnetic braking enhances the accretion speed and hence the radiative shock energy release \citep{commercon_collapse_2011}.
This energy heats the disk and prevents further fragmentation.
The treatment of direct stellar irradiation within our hybrid method can modify this interplay because the disk self-shielding will be captured better than with the FLD method (if the mean-free path is resolved, see Sec.~\ref{sec:pinte}), and also via the launching of magnetic and radiative outflows.
Thus, the hybrid method offers new perspectives for the radiation-magneto-hydrodynamics simulations of massive star formation.

\begin{acknowledgements}
      Part of this work was supported by the CNRS "Programme National de Physique Stellaire" (\emph{PNPS}). The numerical simulations presented here were run on the \emph{CEA} machines \emph{Irfucoast} and \emph{Alfvén}. The visualisation of \ramses{} data has been executed with the \href{https://github.com/nvaytet/osyris}{OSYRIS} python package.
      We thank the referee for constructive comments.
      RMR acknowledges C. Pinte for MCFOST data, and J. Ramsey, R. Kuiper, and S. Fromang for useful discussions.
      RMR finally thanks T. Foglizzo for his fruitful insight on Rayleigh-Taylor instabilities.
\end{acknowledgements}

%
%
\bibliographystyle{aa} 
\bibliography{Zotero} 
%
\begin{appendix} 
\section{Temperature structure with isotropic scattering}
\label{app:isoscat}
   
Figure~\ref{fig:opa_pasc} shows that the extinction opacity is dominated by scattering at high frequency, that is, where stellar irradiation dominates.
Hence, we examine how the gray opacities are modified when including isotropic scattering in the FLD and M1 equations.
It can be shown that taking the scattering into account does not modify the first moment of the RT equation (the conservation of the radiative energy), because the source term involves the radiative energy and its redistribution is the same with and without scattering.

However, the first moment of the RT equation (see Eq.~\ref{eq:ef}) is modified because the coupling between the gas and the radiative flux is enhanced by the scattering. Therefore, the opacity in this equation is the extinction (absorption+scattering) opacity instead of the absorption opacity.
In the gray approximation, the FLD closure relation leads to a Rosseland mean opacity computed with the extinction opacities.
Similarly, the gray version of the flux evolution equation in the M1 module introduces the Planck mean opacity as defined in Eq.~\ref{eq:kp} but with the extinction opacities.

In fact, the inclusion of isotropic scattering has no explicit impact on the radiative energy repartition but it increases the absorption and redistribution of the radiative flux.
To test the behavior of the hybrid approach with isotropic scattering we run the most optically-thick case of the setup from \cite{pascucci_2d_2004}, with $\tau=100$, following the configuration and numerical parameters of section \ref{sec:pasc}.

\begin{figure}
\centering
  \includegraphics[width=9cm]{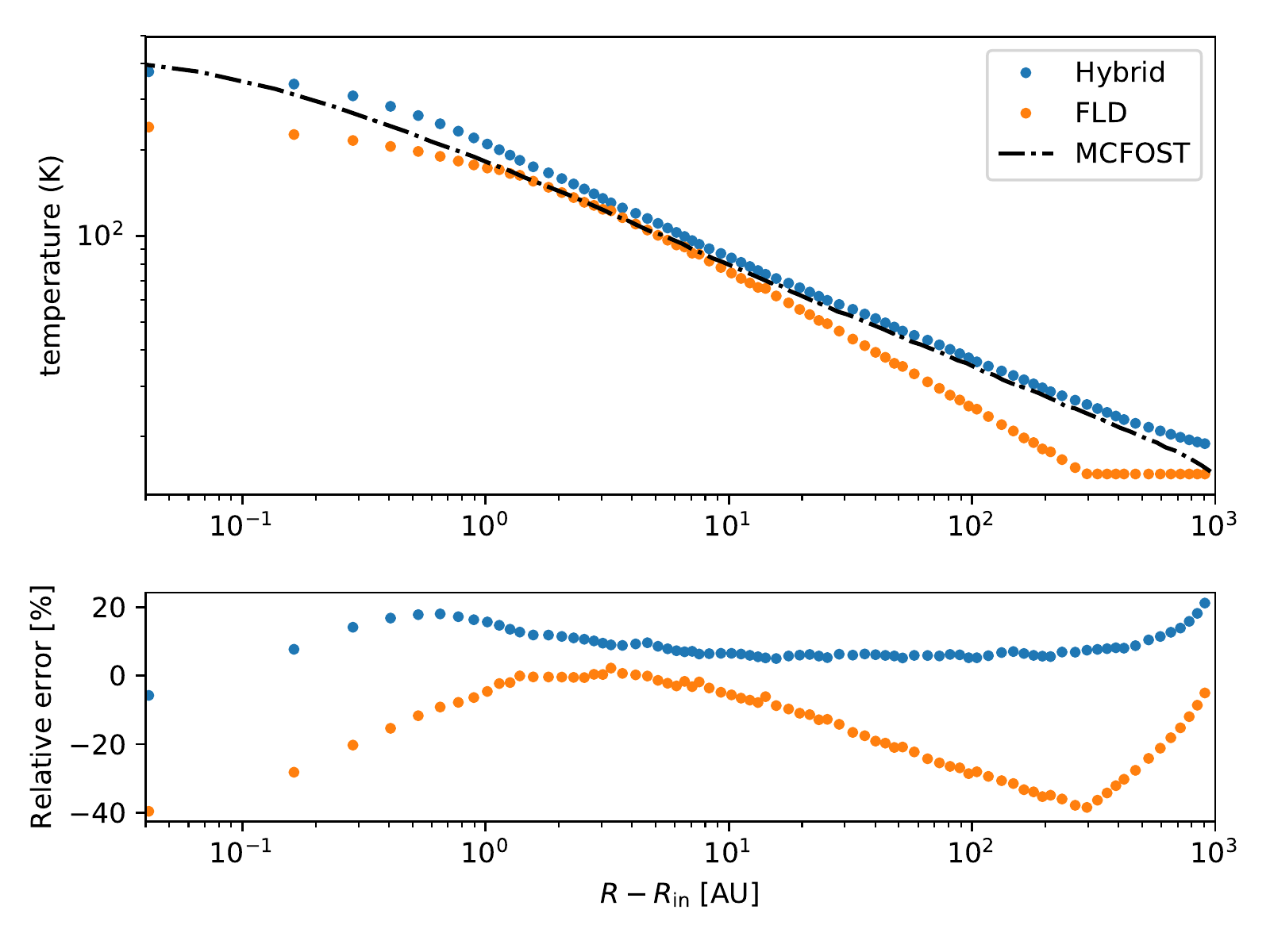}
  \includegraphics[width=9cm]{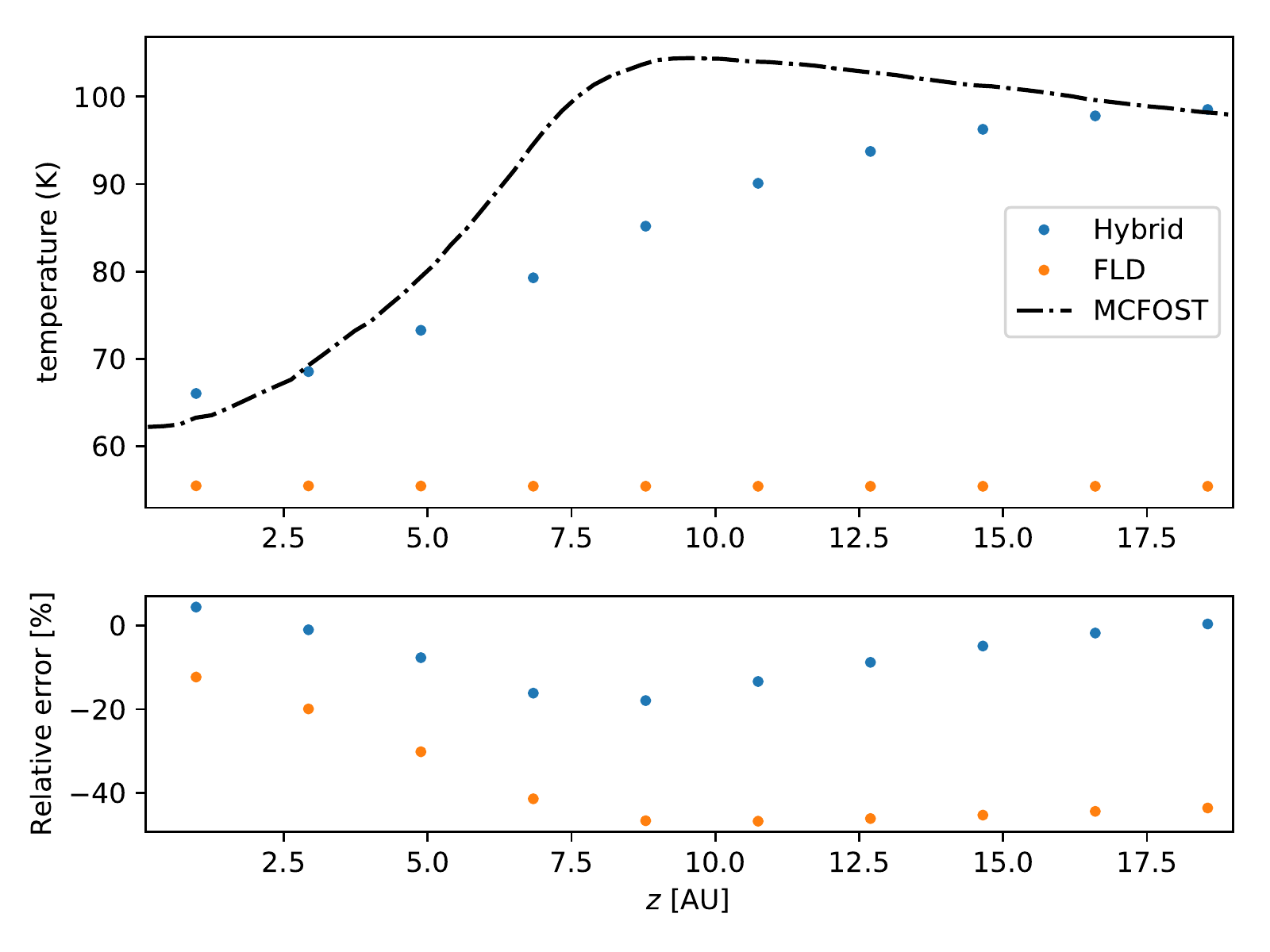}
    \caption{Gas temperature profiles, following the test of \cite{pascucci_2d_2004}. We compare the gas temperature computed using \textsc{MCFOST}, the hybrid method (M1+FLD) and the FLD method alone in \ramses{}, with isotropic scattering. $T_\star = \SI{5800}{K}$ and the integrated optical depth in the disk mid-plane is $\tau=100$. Top: radial profile in the disk mid-plane. Bottom: vertical profile at a disk radius of \SI[mode=text]{20}{AU}.}
    \label{fig:xzTpat2T1_scat}
\end{figure}

The left panel of Fig.~\ref{fig:xzTpat2T1_scat} shows that the temperature in the mid-plane of the disk is well reproduced by the hybrid approach, as compared to the result obtained with \textsc{MCFOST}. The maximal error is ${\approx}20\%$ with the hybrid against ${\approx}40\%$ with the FLD method alone. The right panel of Fig.~\ref{fig:xzTpat2T1_scat} emphasizes that most of the shielding effect in the disk mid-plane is captured by the hybrid approach, unlike for the FLD method.
We compare both panels of Fig.~\ref{fig:xzTpat2T1_scat} with the left panel of Fig.~\ref{fig:x_T_pa_t2_m1vsfldvsmcfost} and right panel of Fig.~\ref{fig:z_T_pa_m1vsfldvsmcfostnoscat}, respectively, as they show the result of the same setup without scattering.
We observe that the temperature structure is not much influenced by the treatment of scattering.
First, as shown in Fig.~\ref{fig:opa_pasc}, scattering only dominates at high-frequencies and therefore mainly at the first interaction of stellar photons with the medium, which corresponds to the disk inner edge. After this interaction, photons are reemitted at the local temperature, where the absorption opacity dominates.
This explains why the temperature at the disk inner edge is ${\sim}\SI{400}{K}$ with isotropic scattering against ${\sim}\SI{350}{K}$ without scattering.
Second, since opacities are frequency-averaged in our hybrid approach, taking isotropic scattering into account does not impact much the Planck and Rosseland mean opacities.
Therefore, we do not consider scattering in the collapse calculations in section \ref{sec:col}.

\section{Performance test}
\label{app:perform}

\begin{figure}
\centering
  \includegraphics[width=9cm]{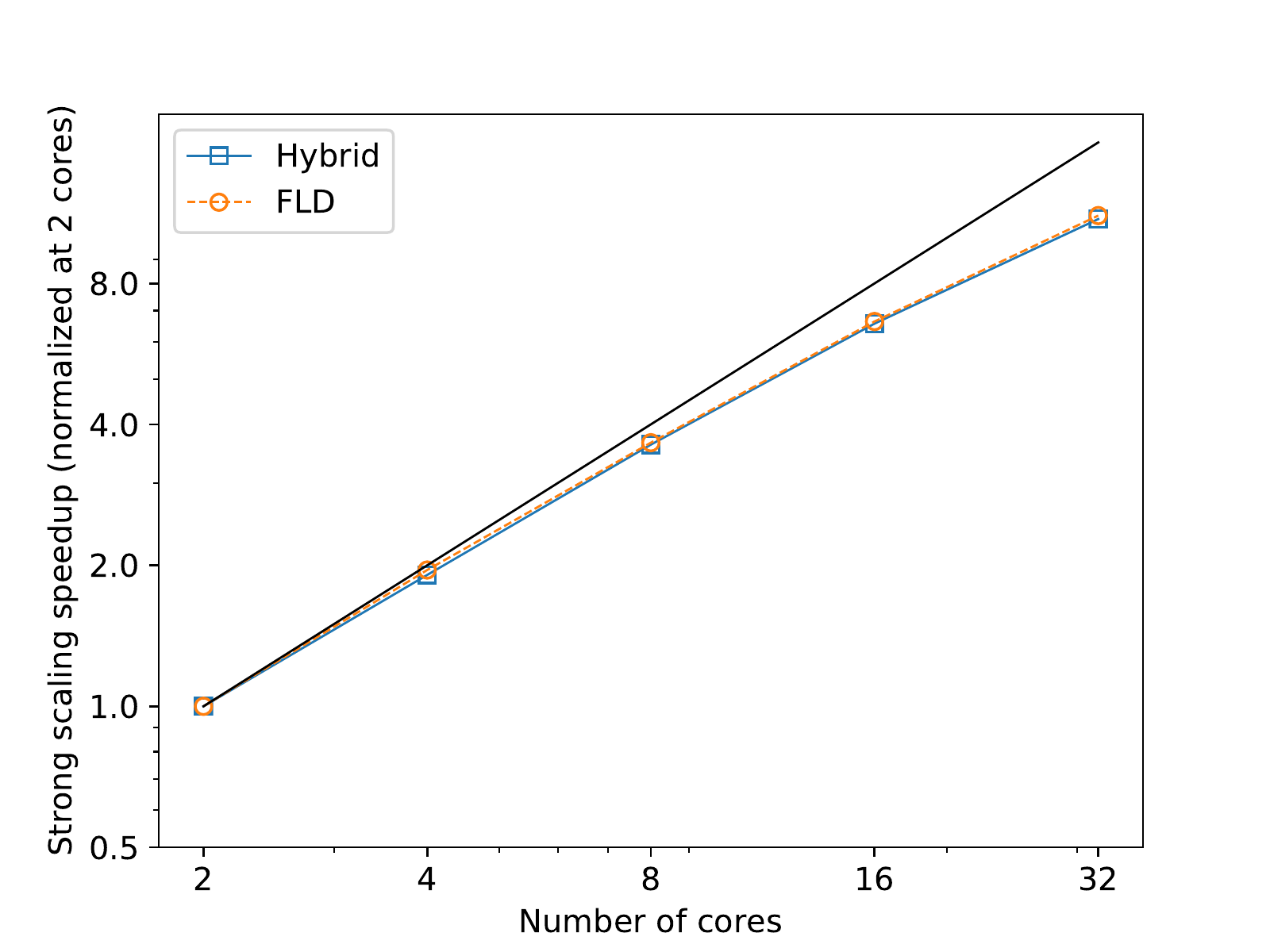}
    \caption{Strong scaling result for $2$ to $32$ cores in the test of \cite{pascucci_2d_2004}, with $\tau=100$, $T_\star = \SI{5800}{K}$. The ideal theoretical speedup is represented by the black line. We compare the strong scaling between the hybrid method (M1+FLD, blue squares) and the FLD method alone (orange circles). We normalize the speedup by that obtained with two cores.}
    \label{fig:str_scaling}
\end{figure}

In this section, we compare the performance of the FLD and the hybrid methods.
First, we run the test from \cite{pascucci_2d_2004}, with $\tau=100$ and $T_\star = \SI{5800}{K}$ to probe the scaling of each method. 
For this test, we choose a grid with two levels of refinement: 8 and 9, which leads to ${\sim} 2 \times 10^7$ cells. 
Figure~\ref{fig:str_scaling} shows the strong scaling results from 2 to 32 cores. 
We can see that the scaling properties of the FLD and the hybrid methods are very similar and close to the theoretical line.
The departure from the theoretical line (speedup of ${\simeq}11$ instead of $16$ for 32 cores) is likely due to the high number of global communications which occur in the FLD conjugate gradient algorithm.

The FLD implementation without hydrodynamics is implicit and therefore is not restricted by the CFL condition in our pure radiative transfer tests, in constrast to the M1 part of our hybrid method.
Therefore, we do not compare the computational time in those tests.
However, we look at the total CPU time in the collapse calculations presented in Sec.~\ref{sec:col}.
The HY run took ${\approx} 5100$ CPU hours, against ${\approx}3900$ for the FLD run, which consists in an additional time of about ${\approx} 30\%$.
The time step in the HY run is first constrained by the M1 CFL condition when the primary sink forms, which is responsible for this difference of computational time: more steps were needed to reach the same physical time.
As mentioned in Sec.~\ref{sec:model}, the FLD modifies the hydrodynamical CFL time step : the sound speed accounts for both thermal and radiative pressures.
It decreases as the radiative pressure (hence energy) increases, while the M1 time step is fixed. 
As the central mass gains mass, its temperature and luminosity generally rise and so does the radiative energy.
Therefore, the FLD time step decreases in both runs, but is still greater than the M1 time step in the HY run, at first.
Then when the outflows are launched, both runs are limited by the FLD time step, because both the radiative energy has become significant and the density is very low in the outflow (hence the modified sound speed increases).
From this time on, the time step is comparable in both runs.
However, the number of iterations in the conjugate gradient is ${\sim} 10\%$ smaller in the HY run than in the FLD run, and so is the elapsed time per time step.

\end{appendix}

\end{document}